\documentclass{article}
\usepackage[utf8]{inputenc}
\usepackage{graphicx}
\usepackage{subcaption}
\usepackage{amsmath}
\usepackage{amssymb}
\usepackage{tablefootnote}
\usepackage{csvsimple}
\usepackage{setspace}
\usepackage{amsthm}
\usepackage{algorithm}
\usepackage{algpseudocode}

\usepackage{bbm}
\usepackage{tikz}
\usepackage{blkarray}
\usepackage{hyperref}
\usepackage[section]{placeins}
\usetikzlibrary{arrows}
\usepackage[
  backend=biber,
  style=apa,
  citestyle=apa
]{biblatex}
\graphicspath{ {./images/}}
\usepackage{geometry}
\title{Non-linear approximations of DSGE models with neural-networks and hard-constraints\thanks{Contact: \href{mailto:emmet.hall-hoffarth@economics.ox.ac.uk}{emmet.hall-hoffarth@economics.ox.ac.uk}. I would like to thank Jeremy Large and Michael McMahon for their thoughtful comments and guidance in the writing of this paper.}}
\author{Emmet Hall-Hoffarth\\University of Oxford}
\date{\today}

\begin{document}

\maketitle

\abstract{Recently a number of papers have suggested using neural-networks in order to approximate policy functions in DSGE models, while avoiding the curse of dimensionality, which for example arises when solving many HANK models, and while preserving non-linearity. One important step of this method is to represent the constraints of the economic model in question in the outputs of the neural-network. I propose, and demonstrate the advantages of, a novel approach to handling these constraints which involves directly constraining the neural-network outputs, such that the economic constraints are satisfied by construction. This is achieved by a combination of re-scaling operations that are differentiable and therefore compatible with the standard gradient-descent approach used when fitting neural-networks. This has a number of attractive properties, and is shown to out-perform the penalty-based approach suggested by the existing literature, which while theoretically sound, can be poorly behaved practice for a number of reasons that I identify.}

\vspace*{\fill}


\newpage

\tableofcontents

\newpage

\section{Introduction}

Recently, there has been an explosion of literature related to the use of neural-networks for the approximation, and even more recently, also the estimation of Dynamic Stochastic General Equilibrium (DSGE) models. This interest comes as a result of the promise of this new approach, which is to alleviate the well-known \textit{curse of dimensionality} attributed to \citeauthor{bellman1957dynamic} (\citeyear{bellman1957dynamic}) that is frequently encountered in the domain of heterogeneous agent models. These benefits are obtained by exploiting the well-documented affinity of neural-networks to high-dimensional and sparse input spaces \parencite{adcock2020deep}. This paper will suggest a modification to the standard approach that has so far been adopted by this literature that addresses some limitations that I have identified, and substantially improves accuracy and performance.

In the process of this approximation, one important step, that is nonetheless not discussed in great detail in the existing literature is the representation of the constraints of the economic model, and the resulting KKT conditions, in the approximating function. For example, an important feature of HANK models is an idiosyncratic borrowing constraint experienced by agents, such that their net wealth cannot fall below some lower bound. This results in agents who are close to or are at the budget constraint having a substantially higher Marginal Propensity to Consume (MPC) than those who are further from the constraint. This dispersion in MPC allows for a pathway connecting inequality and monetary policy, and is important to match many empirical facts \parencite{kaplan2018monetary}. To my knowledge, all of the economics literature related to the application of neural-networks to the approximation of DSGE models implement at least some of these economic constraints by allowing the neural-network to select policies that violate the economic constraints, and then applying a penalty to this violation proportional to either a \textit{Fischer-Burmeister} (FB) function (defined later in (\ref{fischer_burmeister}) in Section \ref{sec:dsge}) in the case of inequality constraints, or simply the (squared) mean deviation in the case of equality constraints. I describe these as \textit{soft-constraint} methods, in contrast to the \textit{hard-constraint} method that I propose in this paper.

While theoretically sound, in practice this soft-constraint approach may be poorly behaved. Since the (squared) FB function is quadratic in its arguments, the penalty becomes extremely small and has a relatively flat gradient in a neighbourhood of the constraint. This means that getting precise approximations is very difficult, as economically meaningful constraint violations may still generate very small losses. Also, since the parameters are learned via gradient descent, the flat gradient means that parameter updates are imprecise. Even when relatively precise estimates are obtained, in this framework there is nothing particular about the actual value of the constraint, so the neural-network choosing an output value exactly on the constraint is a measure-zero event. This may be problematic if the proportion of agents on the constraint is itself a outcome of interest, or if agents' behaviour is expected to change discontinuously at the constraint.

Furthermore, since the learning algorithm is attempting to minimise a loss containing other optimality conditions derived from the first order conditions of the economic model, the optimiser might trade-off the benefit of violating the constraints in terms of improving the other optimality conditions with the essentially arbitrary cost imposed by the penalty functions. For example, the optimiser might choose to consume until the marginal penalty of violating the budget constraint is equal to the marginal utility of consumption. Indeed, the approximated solution should violate the constraints in at least some cases in order for the learning algorithm to be able to get an estimate of the gradient of the loss with respect to the FB function. If, in order to counteract this, a large weight is placed on the FB function, the model may learn to produce outputs that are always far inside the constraints, so as to avoid the penalty entirely. Given these limitations, improving the quality of results relies on either carefully tuning the \textit{hyper-parameters} of the model, for example, the weights placed on penalties related to the various constraints of the model, or training the model for a very long time. Training these neural-networks, while it provides a more general solution, does still take a long time compared to other solution methods, therefore, the necessity of this fine-tuning is an undesirable feature of these techniques.

Perhaps the most salient issue is that when currently approximated policy ($f(X_t, \hat{\theta})$) is used to generate future states ($X_{t+h}$) by simulating the model forward as in \citeauthor{maliar2011solving} (\citeyear{maliar2011solving}), small violations of the constraints will be propagated, which can lead to states being generated that are far away from the true ergodic distribution of states. In some cases this can cause the approximation procedure to breakdown entirely, or require states to frequently be reset to initial values. This is problematic because in the high-dimensional setting where neural-network approximation is likely to have an advantage over other approaches, such endogenous generation of states is necessary to achieve sufficient dimensionality reduction so as to make approximation feasible. Furthermore, there is no guarantee that resetting the states will not result in the same divergence occurring over and over.

In order to address these shortcomings, I propose an alternative approach which involves directly restricting the outputs of the neural-network, so that all of the constraints of the economic model are always satisfied by the output. While it is difficult to present a fully general solution to this problem, I provide methods that are applicable to the majority of currently popular heterogeneous agent models, such as HANK models. This approach resolves all of the previously discussed shortcomings of the soft-constraint approach. It removes the incentive for the learning algorithm to trade off the constraint with other criteria, it ensures that the future generated states are at least feasible, and it allows constraints to bind exactly, if that is an equilibrium outcome. This is achieved through a combination of activation functions that constrain the individual outputs of the neural-network such that the idiosyncratic constraints are satisfied, and contextual re-scaling (more detail in Section \ref{methods}) of the output vector, so that aggregate constraints are also satisfied. The downside of such an approach is that it makes the calculation of gradients significantly more complicated. Fortunately, modern machine-learning software, such as JAX \parencite{frostig2018compiling} can do this automatically and efficiently.

In order to collect more evidence to demonstrate that my hard-constraint approach is superior to penalty-based constraints, in Section \ref{sec:intermediate_cases} this paper will also consider two intermediate constraint regimes, in which some of the constraints are satisfied by construction, while the others are implemented via a penalty. The results, as expected, rest in between the previously discussed cases. Where the constraints are hard, relevant features of the model are faithfully reproduced. Where the constraints are implemented via a penalty, errors accumulate and are quantitatively and qualitatively meaningful. These results show that choosing between constraint implementations, even if they would seem to be equivalent theoretically, can cause substantially different approximations to be obtained. Furthermore, they show that in order to achieve the best level of precision even making a subset of the constraints hard, while simpler to implement, is not sufficient. Instead, the best way to obtain a precise approximation is to treat all constraints with lexicographic preference --- that is to restrict the output of the approximating function to the range of feasible values, before only then attempting to find optimal policies within this subspace.

The remainder of this paper is organised as follows. Section \ref{sec:literature} summarises the relevant literature on using neural-networks to solve DSGE models and HANK models (as used in the application in this paper). Section \ref{methods} outlines the methodological contribution of this paper, which is an algorithm to solve DSGE models with hard-constraints. Section \ref{model} introduces the economic model used in the application in this paper, and explains how the suggested method can be implemented to make the constraints hard for this particular economic model. Section \ref{sec:ml_sol_method} discusses details of the implementation used to produce the results which are provided and discussed in Section \ref{results}. Finally, Section \ref{sec:conclusion} briefly concludes.

\section{Literature} \label{sec:literature}

\subsection{Neural-networks and DSGE} \label{sec:dsge}

The concept of solving control systems using neural-networks has been discussed for a relatively long time, although it has only recently begun to be discussed in the macroeconomics literature. Early papers include \citeauthor{dissanayake1994neural} (\citeyear{dissanayake1994neural}) and \citeauthor{aarts2001neural} (\citeyear{aarts2001neural}). These papers consider the application of neural-networks to solve partial differential equations, a type of problem that is common in physics and engineering that is in many way analogous to solving DSGE models. However, the first paper in this vein in economics was likely \citeauthor{duarte2018machine} (\citeyear{duarte2018machine}), who solves a Lucas style financial asset model with an arbitrary number of assets, as well as a neoclassical growth model, in continuous time, using a neural-network. Subsequent papers which focus specifically on solving DSGE models include \citeauthor{azinovic2022deep} (\citeyear{azinovic2022deep}), \citeauthor{fernandez2020solving} (\citeyear{fernandez2020solving}), and \citeauthor{maliar2021deep} (\citeyear{maliar2021deep}). Particularly relevant to this paper is the \citeauthor{kase2022estimating} (\citeyear{kase2022estimating}), who approximate a 100-agent HANK model, and also show using a simulation that they can recover the true parameters using an estimation technique that combines the particle filter of \citeauthor{fernandez2007estimating} (\citeyear{fernandez2007estimating}) with a likelihood surrogate as in \citeauthor{smith2011estimating} (\citeyear{smith2011estimating}). In this case another neural-network is used as the surrogate.

In general, the process of solving DSGE models involves finding a function $f(X_{t-1}, \epsilon_t; \Gamma)$ of the \textit{states} of the model ($X_{t-1}$), exogenous or structural shocks $\epsilon_t$ and structural parameters $\Gamma$ which maps to choices of agents $Y_t$ that are both optimal in the sense of maximising some objective function and which satisfy constraints on the agents' choices embodied by the Karush-Kuhn-Tucker (KKT) conditions of the model. Since in many interesting cases the function $f$ cannot be found analytically, we instead turn to approximations $\hat{f}(X_{t-1}, \epsilon_t, \Gamma; \theta)$. There are currently two approaches which are commonly used for the approximation of DSGE models. The first, \textit{perturbation} involves approximating $f$ with a Taylor series expansion around the steady state of the model. Especially when this is a first-order approximation this is feasible in a high-dimensional context, in other words, when the dimension of $X_{t-1}$ is large. On the other hand, even if the approximation is a higher order one, by definition this approach is incapable of representing non-linearities, such as kinks or jumps, and the quality of the approximation may be poor if the exogenous shocks are large. The other approach, \textit{projection}, involves estimating the policy function over a number of points in the state-space (known as a grid), and then interpolating between these points, for example with a cubic-spline. This approach is better able to deal with non-linearities, however, the number of grid points required increases quickly with the dimension of the state-space, even if techniques such as \textit{sparse grids} are used \parencite{judd2014smolyak}. Therefore, this technique suffers from the curse of dimensionality. In practice, many macroeconomic researchers use methods based on \citeauthor{reiter2009solving} (\citeyear{reiter2009solving}), which combine projection in idiosyncratic variables with perturbation in aggregates. Still, this is not able to represent non-linear dynamics in the aggregate.

\begin{table}
    \centering
    \begin{tabular}{|c|c|c|c|}
        \hline
        &  & Idiosyncratic & Aggregate \\
        & High-dimensionality & Non-linearity & Non-linearity \\ 
        \hline
        Perturbation & \checkmark & x & x \\
        \hline
        Projection & x & \checkmark & \checkmark \\
        \hline
        \citeauthor{reiter2009solving} (\citeyear{reiter2009solving}) & \checkmark & \checkmark & x \\
        \hline
    \end{tabular}
    \caption{The relative strengths and weaknesses of different approximation techniques}
    \label{nn_advantages}
\end{table}

Taken together, this implies a dichotomy, as discussed in \citeauthor{han2021deepham} (\citeyear{han2021deepham}): none of the existing methods are able to offer a compelling approximation in a both high-dimensional and non-linear (in the aggregate) setting. This gap may be important, in particular, in the growing literature regarding Heterogeneous Agent New Keynesian (HANK) models \parencite{kaplan2018monetary}, which contain both a high-dimensional state space and non-linearities induced by for example an Effective Lower Bound (ELB) on the nominal interest rate. Recently advances in machine learning have pointed towards a potential solution to this problem: neural-networks. The suggestion is simply to use some neural-network functional-form (discussed in more detail later) as the approximating function $\hat{f}(X_{t-1}, \epsilon_t, \Gamma; \theta)$, and then to learn its parameters via Stochastic Gradient Descent (SGD),\footnote{See Appendix \ref{neural_network_background} for a brief description of this.} as is common in machine learning applications. There is a good theoretical justification for doing so thanks to the famous universal approximation of \citeauthor{hornik1989multilayer} (\citeyear{hornik1989multilayer}), which tells us that any function, even if poorly behaved and highly non-linear can be approximated to arbitrary precision by some single-layer, fully connected, and arbitrarily wide feed-forward neural-network. Furthermore, it has been widely documented that neural-networks perform very well in \textit{big data} applications, where high-dimensional inputs are a defining characteristic, and where other approaches therefore suffer from the curse of dimensionality, such as in image recognition and natural language processing \parencite{adcock2020deep,bach2017breaking}. 

Each of the referenced papers concerning the application of neural-networks to solving DSGE models follow the same core approach, which I will outline here. The basic idea is to use a fully-connected, feed-forward neural-network as a global parameterisation of the policy functions that are to be approximated. The inputs of this neural-network are the states of the model which we will represent here with $X_{t-1}$, and the \textit{i.i.d.} exogenous or structural shocks $\epsilon_t$. \citeauthor{kase2022estimating} (\citeyear{kase2022estimating}) also suggest adding the structural parameters $\Gamma$ to the vector of inputs and varying these throughout the training process in order to fit the model for a range of different calibrations at the same time. The outputs of the neural-network are the policy functions $Y_t$. The neural-network contains \textit{trainable parameters} $\theta = \left\{W_i, b_i\right\}_{i=1}^n$ known respectively as weights and biases, which we aim to optimise during the fitting procedure, as well as some fixed \textit{activation functions} $\left\{ \sigma^i \right\}_{i=1}^n$, which are some non-linear vector functions, for example, the ReLu function: $\sigma^i(x)=\max\left\{ 0, x \right\}$. (\ref{eqn:nn_sol}) shows how all of these components are related.

\begin{equation}
    Y_t = \sigma^n\left(W_n \times \sigma^{n-1} (\cdots \sigma^0\left(W_0 \times \left\{ X_{t-1}, \epsilon_t, \Gamma \right\} + b_0\right) \cdots) + b_n\right) \label{eqn:nn_sol}
\end{equation}

The trainable parameters are chosen to minimise a loss function that measures the quality of the fit of the model. In most machine-learning applications this is done by comparing the predictions of the neural-network ($\hat{Y}_t$) to some observed data or \textit{labels} ($Y_t$) that are treated as a ground-truth. However, this application is somewhat different, as the exact optimal policies are not known, since this is what we are trying to solve for. At this point we could create labels via value-function iteration, and essentially interpolate between them using the neural-network, however, this would be essentially equivalent to standard projection methods, and would have all of the same drawbacks, in particular, with respect to the curse of dimensionality. Instead, the quality of the approximation can be evaluated by comparing the predicted policy functions to the optimality conditions of the economic model. This allows for the approximation to remain grid-free. \citeauthor{maliar2021deep} (\citeyear{maliar2021deep}) outline three general ways of constructing the loss function: using the Bellman equation error, the Euler equation error (first order conditions), and in the case of life-cycle models, minus the lifetime utility. In this paper we will focus on the second method using first-order conditions.\footnote{The application in this paper is not a life-cycle model, so the third approach is not relevant. The first approach, using the Bellman equation, despite being more general, unfortunately did not seem to be as well behaved in my own experimentation. I speculate that this because it may be difficult for the optimiser to target the derivative of the outputs of the neural-network ($V^\prime(\cdot)$), although more research is certainly required on the relative strengths and weaknesses of these approaches.} An example of this will given in later in the the application in (\ref{eqn:hank_loss}). This type of loss function has the property that when the loss is exactly zero, the policies satisfy the first order conditions exactly, so they are the exact solution for that given set of states. In practice, the loss will never reach zero because we are dealing with a global approximation. The challenge then, is to make the loss as small as possible, for as many states as possible, in particular, the states that are likely to be generated by the model in equilibrium.

 When the agents face constraints, all of the above referenced papers recommend the same strategy. This is to add a penalty to the loss function calculated using the Fischer-Burmeister function \parencite{fischer1992special} for inequality constraints, as defined in (\ref{fischer_burmeister}), and the squared mean difference for equality constraints. Again, these penalties are exactly zero when the constraints are satisfied by the current policy, and otherwise grow approximately quadratically in the size of the violation of the constraint. The FB function is part of a class of functions known as complementarity functions \parencite{chen2000penalized}. These functions are used as continuously differentiable analogues to KKT conditions (which are required for SGD), and are equal to zero if and only if the KKT conditions are satisfied exactly. Thus, minimising a loss function which includes a complementarity function to zero should also imply that any constraints on that model are also satisfied.

\begin{equation}
    \Psi_{fb}(a, b)^2 = \left(a + b - \sqrt{a^2 + b^2}\right)^2
    \label{fischer_burmeister}
\end{equation}

The parameters of this neural-network are chosen to minimise a loss function via SGD.\footnote{Note that the gradient of the loss with respect to the neural-network parameters is usually calculated exactly via automatic-differentiation and \textit{backpropogation} (See appendix \ref{neural_network_background} for more details) rather than using finite-differences.} In general this means that a \textit{batch} of \textit{inputs} (in this context $\left\{ X_{t-1}, \epsilon_t, \Gamma \right\}$) are randomly sampled, and then the parameters are moved by some small step in the direction of the gradient of the loss with respect to the parameters. In this application the sampling of shocks ($\epsilon_t$) and structural parameters ($\Gamma$) is straightforward, as the distribution of shocks is usually assumed explicitly, and the structural parameters can be uniformly sampled from some predefined range, however, sampling the states ($X_{t-1}$) is somewhat more involved.

Initially, the states (inputs) may be sampled randomly, or simply fixed at some initial condition, however, thereafter the states are generated by simulating forward the state transition of the model using the current approximation of the policy function. In this way the ergodic set of the equilibrium and the policy functions are learned simultaneously. This is particularly beneficial because the set of states that are actually visited in equilibrium is usually a very small subset of the hypercube spanned by the full state-space \parencite{maliar2011solving}, and this difference is particularly pronounced when the dimensionality of the state-space is high. Another key difference to other methods in the vein of \citeauthor{reiter2009solving} (\citeyear{reiter2009solving}), that allows for strong dimensionality reduction is the fact that the neural-network represents a global functional form for the approximation over the entire state space, rather than an interpolation between knots. The lack of necessity for discretization means that when using the neural-network approach the total dimension of the inputs grows much more slowly in the number of states of the model.

Nonetheless, this procedure is not without drawbacks. In particular, as noted by \citeauthor{azinovic2022deep} (\citeyear{azinovic2022deep}), at the beginning of the estimation procedure, when the quality of the policy function approximation is poor, following the implied state-transition may lead to very unrealistic states, or even impossible states, if the constraints are violated by that policy. This can slow down convergence significantly, and in some cases cause the approximation procedure to break down entirely. Furthermore, the lack of knots means that we do not get an exact policy for any state, but rather an approximation for all states. This means that we have to rely heavily on the ability of the neural-network to approximate well over the high-dimensional input space in order to get precise estimates. Fortunately, it is possible to measure the quality of the approximation via the loss function, and as demonstrated in for example \citeauthor{azinovic2022deep} (\citeyear{azinovic2022deep}) the approach is highly scaleable, such that additional computational resources can be added to improve the quality of the approximation to an acceptable standard.

\subsection{Limitations of Penalty-Based Constraints}

Using a penalty for constraint violation is not specific to NN-based solution methods. For example, in \citeauthor{achdou2022income} (\citeyear{achdou2022income}) penalties are applied for violations of constraints in their continuous-time solution method. Indeed, most approaches to solving dynamic models involve penalties for constraint violation. When the model is solved over some form of grid of points this is an appropriate implementation because it is computationally efficient, straightforward to implement, and can be made arbitrarily accurate by continued iteration. However, what seems to be particularly problematic in practice is the combination of penalties with endogenous sampling of states as in \citeauthor{maliar2011solving} (\citeyear{maliar2011solving}).\footnote{Note that these states are not endogenous in the same sense as the endogenous grid points of \citeauthor{carroll2006method} (\citeyear{carroll2006method}). Those are still grid points that are chosen ex-ante, which are endogenous in the sense that they are treated as end-of-period states. In this case on the other hand the states are truly endogenous in the sense that they change throughout the approximation procedure and depend directly on the current policy function.}. Usually, when this forward-simulation is used the states are simulated forward many periods at a time.\footnote{For example \citeauthor{kase2022estimating} (\citeyear{kase2022estimating}) simulate forward 20 periods between each parameter update, while \citeauthor{maliar2021deep} (\citeyear{maliar2021deep}) simulate 10 periods forward.} This is done in order to make sure that states are approximately uncorrelated \parencite{maliar2021deep}, and also in the case of the extended neural-network of \citeauthor{kase2022estimating} (\citeyear{kase2022estimating}) that the states represent a true draw from the ergodic distribution under freshly redrawn structural parameters. As a result, small violations of the constraints such as market clearing conditions have the potential to compound between draws of states. 

\begin{figure}
    \centering
    \captionsetup{width=0.8\textwidth}
    \includegraphics[width=0.8\textwidth]{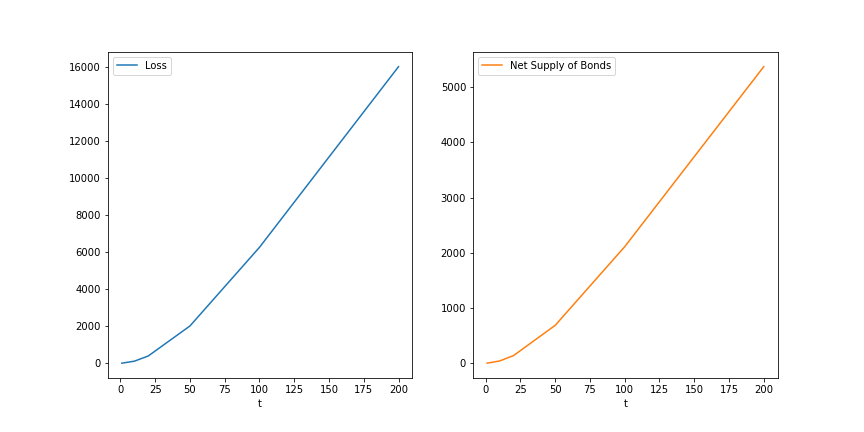}
    \caption{Overall loss (left) and net supply of bonds (right) obtained by simulating forward the model (Section \ref{model}) with randomised initial parameters when constraints are not enforced (soft-constraint method). Over 200 iterations the net supply of bonds grows exponentially away from its equilibrium value (0), and as a result so too does the overall loss.}
    \label{fig:soft_init_divergence}
\end{figure}

For example, consider a model in which the initial policy estimates imply an artificially high interest rate, and a net supply of bonds is greater than zero (the equilibrium value). Then even though the penalty term and subsequent parameter update implies that the net bond positions should reduce, after simulating forward a number of periods from the initial state the return on these net-excess bonds would be so large that the model would have already moved far away from equilibrium. Converging back down to smaller bond positions from here could be slow since we are so far from realistic values the gradient may be very large and not very informative. In principle, if the system is globally convergent to a single equilibrium then it should still be possible to find it despite this compounding, however, in practice, it is common to run into computational issues, if for example bond positions increase to above the largest values that can be represented or consumption goes to zero (which implies division by zero in the FoCs in the case of log-utility). Figure \ref{fig:soft_init_divergence} clearly demonstrates this divergence by simulating forwards a model with randomised initial parameters for 200 periods, in the case where constraints are not strongly enforced. 

As a result of this, in my own experimentation with implementing methods based on the soft-constraint approach I found that a number of modifications seem to be necessary in order to keep states within the set of feasible values and to render the approximation procedure stable. Firstly, the number of forward-simulation between updates must be kept low, at least early in the approximation procedure. This can then be relaxed by gradually increasing the number of forward simulations between parameter updates as long as the constraint-related loss components stay satisfactorily small. Secondly, a larger learning rate can be used, in order to counteract the divergent nature of initial solutions. This learning rate can then subsequently be lowered as training iterations go on. Furthermore, one can train for multiple parameter updates with current estimated states by repeatedly drawing random sub-batches, before performing the next forward simulation.\footnote{This is the approach in for example \citeauthor{azinovic2022deep} (\citeyear{azinovic2022deep}), however, it is not necessary in the hard-constraint approach, which is stable even if forward simulation is performed after every parameter update} This helps to balance out the how often forward simulation takes place relative to parameter updates. 

Finally, despite all of these adjustment it is still the case that for some initial conditions the approximation procedure will still break-down, resulting in incredibly large losses, a corner solution, or NaNs (invalid policies that cause division by zero). In these cases the gradient is ignored, and the states have to be reset to initial values. When approximation diverges it is not possible to for example simply set the loss to a large or infinite value because doing so would not result in an informative gradient that points the optimiser back towards reasonable parameters. Since there is no obvious relationship between individual neural-network parameters and outputs there is no straightforward way to manually adjust parameters to move away from these extreme values; this is exactly what we hope the gradient will tell us. Furthermore, repeatedly resetting states is undesirable because if nothing else changes there is no guarantee that the optimiser will not end up at the same corner solution time and time again.

If all of these accommodations are made it is possible to achieve convergence in policy estimates under soft-constraints, however, these accommodations and refinements would not be necessary if the approximated policies could somehow be manipulated so as to always satisfy the constraints of the model. Furthermore, as I will demonstrate later, doing so also makes the resulting approximation of policy functions significantly more accurate, especially close to those constraints.

\subsection{HANK}

One clear use-case for these machine-learning techniques is given by HANK models, such as the model proposed by \citeauthor{kaplan2018monetary} (\citeyear{kaplan2018monetary}). These models combine price rigidity and imperfect competition from standard RANK models with the incomplete markets and heterogeneity of \citeauthor{krusell1998income} (\citeyear{krusell1998income}) type models in order to provide a role for inequality in the analysis of monetary policy \parencite{acharya2020optimal}. The primary goal of HANK models is to achieve more realistic Marginal Propensities to Consume (MPCs), and more realistic monetary policy transmission, when compared with their RANK counterparts. HANK models explicitly take into account the fact that the way that agents interact with financial markets may vary significantly depending on the wealth of agents, and allow for the extent to which agents are constrained to be determined endogenously.

Because heterogeneity is a core feature of these models, they tend to involve high dimensional state-spaces, especially when there are a number of different asset classes modelled. One approach to dealing with this is to assume a limited form of heterogeneity. For example Two-Agent New-Keynesian (TANK) models exogenously assume a mass of agents are financially excluded and therefore must consume all of their current wealth \parencite{gali2007understanding}. This captures much of the intuition of HANK models, but since the types of agents are exogenous it fails to capture the fact that severe shocks may meaningfully impact the degree of inequality in the economy and hence aggregate dynamics though this channel. Other papers have developed methodological advances that make solving models with rich and endogenous heterogeneity more feasible. For example, \citeauthor{bayer2018solving} (\citeyear{bayer2018solving}) and  \citeauthor{auclert2021using} (\citeyear{auclert2021using}), and \citeauthor{ahn2018inequality} (\citeyear{ahn2018inequality}) augment the methodology of \citeauthor{reiter2009solving} (\citeyear{reiter2009solving}) using various techniques to achieve greater amounts of dimensionality reduction and thus render larger models possible to solve. However, these methods all rely on linearisation in the response to aggregate shocks. Therefore, the primary benefit of the neural-network solution method in comparison to these more standard methods is the ability to capture non-linear aggregate dynamics.

\section{Methodology} \label{methods}

The primary methodological contribution of this paper is the utilisation of \textit{hard-constraints} on neural-network outputs, in order to ensure that policy functions are restricted to the space of functions which satisfy all model constraints. Unfortunately, it would be very difficult or impossible to specify a solution of this nature that is completely general. However, it is possible to specify solutions to a number of important cases that cover most heterogeneous agent models used in practice. To fix ideas, consider the standard state-transition equation, where a neural-network is used to approximate the policy function:

\begin{align}
    Y_t =& \sigma^n\left(W_n \cdot \sigma^{n-1} (\cdots \sigma^0\left(W_0 \cdot \left\{ X_{t-1}, \epsilon_t, \Gamma \right\} + b_0\right) \cdots) + b_n\right) \\
    X_t =& S(X_{t-1}, \epsilon_t, Y_t; \Gamma) \\
    \text{s.t. } & g(X_t, Y_t) \geq 0
\end{align}

Then, the suggestion is to add a scaling function $h$ to the neural-network outputs in order to ensure that the constraints $g$ are always satisfied, regardless of the output of the neural-network. We can write this as:

\begin{align}
    Z_t =& \sigma^n\left(W_n \cdot \sigma^{n-1} (\cdots \sigma^0\left(W_0 \cdot \left\{ X_{t-1}, \epsilon_t, \Gamma \right\} + b_0\right) \cdots) + b_n\right) \\
    Y_t =& h\left( Z_t \right)
\end{align}

This can be thought of as a separate operation after the neural-network outputs, or simply as a complex type of output layer. There are a number of common types of constraints for which a straightforward re-scaling can be applied, which will now be discussed briefly. If the constraint is of the form $a < Y_t$ then the output can be constructed as $Y_t = a + softplus(Z_t)$ or $Y_t = a + exp(Z_t)$, where $Z_t$ is the output of the last layer in the neural-network, and $softplus$, which is a common activation function used in machine-learning applications, is defined as:

\begin{equation*}
    softplus(x) = log(1 + exp(x))
\end{equation*}

Likewise, if the constraint is of the form $Y_t < b$ the output can be constructed as $Y_t = b - softplus(X_t)$ or $Y_t = b - exp(X_t)$. Furthermore, if the constraint is of the form $a < Y_t < b$ the output can be constructed as $Y_t = a + \left( b - a \right) \cdot sigmoid(Z_t)$, where $sigmoid$ is defined as\footnote{However, in order to avoid a \textit{vanishing gradient problem}, it may be better to simply allow initial values outside of these constraints, and then clip them back inside, see Section \ref{sec:constraint_satisfaction}.}:

\begin{equation*}
    sigmoid(x) = \frac{1}{1+exp(-x)}
\end{equation*}

While these individual constraints are straightforward, more difficult problems can arise when $Y_t$ is a vector, and the constraints imply some relationship between its elements. A prominent, and quite general example is the constraint $a^i < Y^i_t < b^i; \sum_i Y^i_t = C$ for $C \in \left(\sum_i a^i, \sum_i b^i\right)$. In the case where $a^i = 0; b^i = C = 1 \forall i$, then this is a well known problem where the outputs can be interpreted as predicted probabilities of various mutually exclusive categories. In machine learning, predictions for these kinds of \textit{multi-label classification} problems are handled by the \textit{softmax} function, which is defined as:

\begin{equation*}
    softmax(x) = \frac{exp(x)}{\sum_i exp(x^i) }
\end{equation*}

Another solution, which will be used in parts of Algorithms \ref{alg:rescale1} and \ref{alg:rescale2} is a affine transformation that shifts by the minimum value and scales by the sum: $Y_t =\frac{Z_t - \min_i Z^i_t}{\sum_i Z_t - \min_i Z^i_t}$. If $C \neq 1$, then $C$ can simply be multiplied by these functions to change their sum appropriately. However, when $a^i$, $b^i$ and $C$ are unrestricted this problem is significantly more difficult, and no obvious solution exits. A naive approach might be to first re-scale outputs so that they fall between the upper and lower bounds, and then re-scale again so that the sum of outputs equals C. These steps are summarised in the following two equations:

\begin{align}
    z^\prime &= \frac{(b - a) x}{\|x\|} + a \\
    z &= C \frac{z^\prime}{\|z^\prime\|}
\end{align}

However, the problem with this is that is that re-scaling done to satisfy the lower and upper bounds $a^i$ and $b^i$ may change the sum of outputs, and thus cause the summation constraint to be violated, and vice versa. Of course, we could always use an iterative algorithm to find a valid solution, however, this is not appropriate in this application, because the outputs of such an algorithm would not bet differentiable with respect to the neural-network parameters, at least, not using backpropogation and automatic differentiation. Furthermore, performing such an iteration within every policy evaluation would slow down computation substantially, perhaps enough to render approximation infeasible.

\begin{algorithm}
\caption{Re-scaling Algorithm}
\label{alg:rescale1}
\begin{algorithmic}
    \Require $x \in \mathbb{R}_+^k, a \in \mathbb{R}_+^k, b \in \mathbb{R}_+^k, C \in \mathbb{R}_+, a^i < b^i \forall i, \sum_i a^i < C < \sum_i b^i$
    \Ensure $w \in R_+^k : a^i < w^i < b^i \forall i, \sum_i w^i = C, w^i \propto x^i$
    \State $z^\prime \gets (b - a) \frac{x}{\sum_i x^i} + a$
    \State $z \gets C * \frac{z^\prime}{\sum_i z^{\prime i}}$
    \State $A \gets \sum_{z^i < a^i} \left( a^i - z^i \right)$
    \For {$i \in \left\{ 0, ..., L \right\}$}
        \State $\hat{a^i} \gets \max \left\{0, w^i - a^i \right\}$
    \EndFor
    \For {$i \in \left\{ 0, ..., L \right\}$}
        \State $w^i \gets \min \left\{ a^i, z^i - \frac{\hat{a}}{\sum_i \hat{a}} A \right\}$
    \EndFor
    \State $B \gets \sum_{w^i > b^i} (w^i - b^i)$
    \For {$i \in \left\{ 0, ..., L \right\}$}
        \State $\hat{b^i} \gets \max \left\{0, b^i - w^i  \right\}$
    \EndFor
    \For {$i \in \left\{ 0, ..., L \right\}$}
        \State $w^i \gets \min \left\{b^i, w^i + \frac{\hat{b}}{\sum_i \hat{b}} B \right\}$
    \EndFor
\end{algorithmic}
\end{algorithm}

Instead, Algorithm \ref{alg:rescale1} describes a function that solves this problem using only vectorised operations,\footnote{The algorithm shows many steps as for-loops over $i$ but these should be understood to be, and are actually in the implementation, vectorized operations.} which is therefore differentiable and parallelisable. The logic of this algorithm is as follows. The first step is to re-scale each $x^i$ to be within the upper and lower bounds. We then take the result of this and re-scale it again so that the sum of the elements is $C$. However, as previously noted, this second re-scaling may undo the first, in the sense that it can cause some of the elements of the resulting vector to be outside the lower and upper bounds. Therefore, we introduce a final step, in which we find the total amount that would have to be added to or subtracted from each element outside the constraint in order to move it exactly to the constraint, and then distribute this sum onto the unconstrained elements by an amount that is  proportional to their distance to the bounds, such that the total sum of the vector remains unchanged. The weight is distributed proportionally to the distance to the boundaries in order to minimise the probability that any new elements are pushed over their own boundary. We do this operation first for the lower bound $a$ and then for the upper bound $b$, or vice versa, depending on the context. The difference is that the bounds will be able to bind only for the operation that is done last, for the other constraint, the outputs will satisfy the constraint strongly. Note that although these operations force the outputs within the constraints, it does not force the constraints to bind. In other words, it is still possible that the output of this algorithm can be strictly interior relative to the constraints, if this is the equilibrium outcome.

This type of constraint may seem quite arbitrary, but it is actually very general. In particular, it will occur in any heterogeneous agent model with a discrete number of agents and incomplete markets. In this context the idiosyncratic constraints $a^i$ and $b^i$ relate to the agents' choice of expenditure --- that is the net of total income and the constrained resource --- which must be strictly positive and also feasible with respect to the borrowing constraint. Notice that this concerns expenditure, not consumption, as it may also include investments in other assets should they be present in the model. The aggregate constraint $C$ relates to the market clearing condition of the economy. Even if there are multiple goods or assets agents can purchase, at some point we will need to limit the sum of their expenditure in order to ensure market clearing due to Walrus' law. See Section \ref{sec:constraint_satisfaction} for a particular example of how this is applied. 

\begin{algorithm}
\caption{Alternative Re-scaling Algorithm}
\label{alg:rescale2}
\begin{algorithmic}
    \Require $x \in \mathbb{R}_+^k, a \in \mathbb{R}_+^k, b \in \mathbb{R}_+^k, C \in \mathbb{R}_+$, $a^i < b^i \forall i, \sum_i a^i < C < \sum_i b^i$
    \Ensure $w \in R_+^k : a^i < w^i < b^i \forall i, \sum_i w^i = C, w^i \propto x^i$
    \State $z^\prime \gets (b - a) \frac{x}{\sum_i x^i} + a$
    \State $z \gets C \cdot \frac{z^\prime}{\sum_i z^{\prime i}}$
    \State $T \gets \sum_i \max_i \left(z^i - b^i, 0\right) + \sum_i \min_i \left(z^i - a^i, 0\right)$
    \If {$T \geq 0$}
        \For {$i \in \left\{ 0, ..., L \right\}$}
            \State $r^i \gets \max_i \left( b^i - z^i, 0 \right)$
        \EndFor
    \Else
        \For {$i \in \left\{ 0, ..., L \right\}$}
            \State $r^i \gets \max_i \left( z^i - a^i, 0 \right)$
        \EndFor
    \EndIf
    \State $\tilde{r} \gets \frac{r}{\sum_i r^i}$
    \State $w \gets \min\left(\max\left(z, a\right), b\right) + \tilde{r}T$
\end{algorithmic}
\end{algorithm}

An alternative approach is given in Algorithm \ref{alg:rescale2}. This function essentially wraps the final two steps of Algorithm \ref{alg:rescale1} into one. As a result it has the property that both the lower and upper bounds can bind in the output. This may or may not be desirable depending on the context. The purpose of including this is to demonstrate that there is not necessarily a unique solution to this problem, there may be multiple each with their own trade-offs. However, neither of these algorithms are not completely robust because they can fail in a particular edge case. In particular, if $A$ or $B$ are particularly large, or $\sum\hat{a}$ or $\sum\hat{b}$ are particularly low, then the weight that is dispersed to the unconstrained elements may be so large that it pushes those elements over their own bounds. In other words, they may be increased or decreased by a proportion of their distance to the bound that is greater than one. These conditions correspond to the case where there is very large wealth inequality between agents, and the proportion of agents who are constrained by the budget constraint is high, such that that the remaining agents save in order to satisfy the market-clearing condition that their consumption is negative. Fortunately, this was not encountered in the application in this paper, because under all of the allowed combinations of parameters the equilibrium does not touch this corner case.

If another application involves this type of extreme inequality as an equilibrium outcome, there is one further modification that can be made to mitigate the possibility of generating values outside the bounds. Suppose for the following that it is the upper bound $b$ that we are concerned about breaking. Since for any solution to exist we have assumed that $\sum_i a^i < C < \sum_ib^i$, then for any given values of $a$, $b$, and $C$ there is a maximum number of elements of the output vector that can be at the upper bound, such that it is still possible for the vector to sum to $C$. We can find this amount in a vectorised fashion by creating a $(2\times K)$ vector of $a$, and $b$, sorting by $b$ decreasing, and finding the maximum index $i^*$ such that the sum of the cumulative sum of the sorted vector $b$ up to $i^*$ and the cumulative sum of the sorted vector $a$ after $i^*$ is less than $C$. We can then use this to make sure the output vector falls within the bounds by shifting $w^i$ appropriately. In particular, we can sort $z^i$ decreasing, and then find $z^{i*}$. We then shift the vector $z^i$ by subtracting $\bar{z} = \frac{1}{\frac{b^{i*}}{C} L - 1} \left(\frac{b^{i*} \sum{z^i}}{C} - z^{i*} \right)$ (see Appendix \ref{appendix:zbar} for proof). Essentially, this ensures that the resulting $z^{i*} = b^{i*}$ so all $z^i$ which are less fall strictly within the upper bound, such that less than the maximum amount of constrained elements that can feasibly hit the upper bound $b^i$ actually do in the resulting output.

\section{Model} \label{model}

\subsection{Households}

In order to demonstrate how this constraint regimes can be implemented, I will introduce a HANK model, based on that in \citeauthor{kase2022estimating} (\citeyear{kase2022estimating}) and explain how to constrain the outputs of this model. The model contains a discrete number $L$ of heterogeneous agents (which is set to $100$ in all of the applications), indexed by $i$ who maximise a utility function $U$ subject to idiosyncratic shocks and budget constraints. The households' optimisation problem can be summarised as:

\begin{align}
    \max_{\{c^i_t, h^i_t\}} U^i =& \mathbb{E}_0 \left[\sum_{t=0}^\infty \beta^t exp(\Psi_t) \left(\frac{1}{1-\sigma} (c^i_{t} - h C_{t-1})^{(1 - \sigma)} - \chi \frac{1}{1+\eta} \left({h^i_t}^{1+\eta}\right)\right)\right] \\ 
    &\text{s.t.} \nonumber \\
    &c^i_t + b^i_t = W_t s^i_t h^i_t + Div^i_t + \frac{R_{t-1}}{\Pi_t} b^i_{t-1} \equiv \omega^i_t \label{budget_constraint} \\ 
    &b^i_t \geq \underline{B} \label{borrowing_constraint} \\
    &c^i_t > 0 \label{consumption_constraint} \\
    &h^i_t > 0 \label{labour_constraint} \\
    &\Psi_t = exp\left(\rho_\Psi log(\Psi_{t-1}) + \sigma_\Psi \epsilon^\Psi_t \right) \nonumber \\
    &s^i_t = exp\left(\rho_s log(s^i_{t-1}) + \sigma_s \epsilon^{s, i}_t \right) \nonumber
\end{align}

All shocks denoted $\epsilon$ are distributed $iid$ $N(0, 1)$. $\Psi_t$ is an aggregate preference shock, while $s^i_t$ is an idiosyncratic labour productivity shock. In the implementation $s^i_t$ are re-scaled to have a mean of one in all states, so that these idiosyncratic shocks have no direct aggregate effect. However, we do not assume \citeauthor{greenwood1988investment} (\citeyear{greenwood1988investment}) preferences, so it is possible that the distribution of $s^i_t$ can indeed have aggregate implications, through agents' differential responses to income risk depending on their level of wealth. The resulting first order conditions for the household are:

\begin{align}
    1 - \mu^i_t =& \beta R_t \mathbb{E}_t \left[ \frac{exp(\Psi_{t+1})}{exp(\Psi_t)} \left( \frac{\lambda^i_t}{\lambda^i_{t+1}} \right)^\sigma \frac{1}{\Pi_{t+1}} \right] \label{euler_equation} \\
    \lambda^i_t \equiv& \left(c^i_t - hC_{t-1}\right)^{-\sigma} = \frac{\chi \left(h^i_t\right)^\eta}{s^i_tW_t} \label{labour_supply_equation} \\
    \mu^i_t \left(b^i_t - \underbar{B}\right) =& 0, b^i_t \geq \underbar{B}, \mu^i_t \geq 0 \label{kkt}
\end{align}

Where (\ref{euler_equation}) is the households' Euler equation, (\ref{labour_supply_equation}) is the households' labour supply equation, and (\ref{kkt}) are the KKT conditions associated with the borrowing constraint. 

\subsection{Firms}

The firms are standard for a New Keynesian setting. A continuum of identical monopolistically competitive firms produce intermediate goods subject to a quadratic \citeauthor{rotemberg1987new} (\citeyear{rotemberg1987new}) price-adjustment cost $\phi$. This results in a New Keynesian Phillips Curve (NKPC):

\begin{equation}
    \phi \left( \frac{\Pi_t}{\Pi} - 1 \right) = (1 - \epsilon) + \epsilon MC_t + \beta \phi \mathbb{E}_t \left[ \frac{\Pi_{t+1}}{R_t} \left( \frac{\Pi_{t+1}}{\Pi} - 1 \right) \frac{\Pi_{t+1}}{\Pi} \frac{Y_{t+1}}{Y_t} \right] \label{nkpc_equation}
\end{equation}

Where $MC_t$ are real marginal costs. The firms are owned by the households, and distribute profits evenly. Therefore $Div^i_t = Div_t = Y_t - W_t N_t$. However, we assume that the firm discounts future profits as if they were owned by an unconstrained agent, such that their discount factor is $\frac{\Pi_{t+1}}{R_t}$.\footnote{This is done primarily to improve computational performance, however, I have tested a version instead using stochastic discount factor (average relative marginal utility of each agent) and the results were not meaningfully different.} In equilibrium, it is assumed that intermediate firms make positive profits and therefore we impose $W_t < \frac{Y_t}{N_t}$ such that $Div_t > 0$. Final goods are produced by a representative competitive firm subject to an AR(1) TFP process $A_t = exp\left(\rho_A log\left(A_{t-1}\right) + \sigma_A \epsilon^A_t\right)$. Labour is hired in a competitive market such that:

\begin{equation}
    MC_t = W_t / A_t
\end{equation}

\subsection{Monetary Authority}

The monetary authority sets the nominal interest rate $R_t$ according to a dual-mandate Taylor rule, with persistence parameter $\rho_R$ and subject to a monetary policy shock $\epsilon^{mp}_t$:

\begin{align}
    R_t &= \left(R_{t-1}\right)^{\rho_R} \left( R \left( \frac{\Pi_t}{\Pi} \right)^{\theta_\Pi} \left( \frac{Y_t}{Y} \right)^{\theta_Y} \right)^{\left(1 - \rho_R\right)} exp(\sigma_{mp}\epsilon^{mp}_t) \label{mon_auth_rule} \\
    R =& \frac{\Pi}{\beta}
\end{align}

\subsection{Resource Constraints, Market Clearing, and Equilibrium}

Finally, there are two resource constraints in the economy. The first stipulates that bonds are in zero net-supply, and the second is an output constraint that requires that production of final goods is equal to total effective labour input which is in turn equal to demand for final goods.

\begin{align}
    \frac{1}{L}\sum_{i=1}^L b^i_t =& 0 \label{resource_constraint} \\
    \frac{1}{L}\sum_{i=1}^L s^i_t h^i_t =& N_t \label{labour_mkt_clearing} \\ 
    Y_t = A_t N_t =& \frac{1}{L}\sum_{i=1}^L c^i_t \label{output_constraint}
\end{align}

Note that if (\ref{output_constraint}) and (\ref{budget_constraint}) hold for $i$, $t$, then (\ref{resource_constraint}) follows by construction, as a result of Walrus' law.

\subsection{Equilibrium}

A Functional Rational Expectations Equilibrium in this model consists of a set of idiosyncratic policy functions $\{c^i_t, h^i_t, \mu^i_t\}_{i=1}^L$, and prices $\{\pi_t, W_t, R_t\}$ such that for all states $\{s_{t-1}^i, A_{t-1}, \Psi_{t-1},\\ b^i_{t-1}, C_{t-1}, R_{t-1}\}_{i=1}^L$ and shocks $\{ \epsilon_t^{s, i}, \epsilon_t^A, \epsilon_t^\Psi, \epsilon_t^{mp} \}_{i=1}^L$ the idiosyncratic policies satisfy the households' optimality conditions (\ref{euler_equation}), (\ref{labour_supply_equation}) and (\ref{kkt}) taking prices as given, the prices satisfy the NKPC (\ref{nkpc_equation}) and the monetary policy rule (\ref{mon_auth_rule}), and the markets for bonds (\ref{resource_constraint}), labour (\ref{labour_mkt_clearing}) and goods clear (\ref{output_constraint}).

\section{Machine Learning Solution Method} \label{sec:ml_sol_method}

\subsection{Loss Function}

In order to solve the model, we will approximate the optimal policy functions using a neural-network, as described previously. The neural-network used will be fully-connected and feed forward, with a depth of 5 layers, and 128 perceptrons per layer. The inputs of the neural-network, as in \citeauthor{kase2022estimating} (\citeyear{kase2022estimating}) will consist of the states of the model ($\left\{ \Psi_{t-1}, s^i_{t-1}, b^i_{t-1}, C_{t-1}, R_{t-1} \right\}_{i=1}^L$), the exogenous shocks ($\left\{ \epsilon^{s,i}_t, \epsilon^{\Psi}_t, \epsilon^{a}_t, \epsilon^{mp}_t \right\}_{i=1}^L$), and the structural parameters of the model, which will be sampled uniformly over a predefined range. The neural-network will output two aggregate policies $\Pi_t$ and $W_t$, and three idiosyncratic policies: $\left\{ \tilde{c}^i_t, h^i_t, \tilde{\mu}^i_t \right\}$. The tilde implies that some of these outputs will be further transformed before they can be interpreted as policy functions, as described in Section \ref{output_constraints}. In practice when approximating the idiosyncratic policies we feed the inputs into a separate neural-network that in addition to the aggregate inputs also contains the idiosyncratic states $s^i_{t-1}, b^i_{t-1}, \epsilon^{s,i}_t$ of each agent. Thus the aggregate policy neural-network has $3L + 31$ inputs and $2$ outputs, and the idiosyncratic policy neural-network has $3L + 34$ inputs and $3$ outputs.

The loss function that the model aims to minimise consists of the product of the first order condition errors for a given state and two independent draws of future shocks (indexed here by $j$) as in \citeauthor{maliar2021deep} (\citeyear{maliar2021deep}). This is done instead of squaring the loss resulting from a single draw of shocks in order to allow for the SGD process to, over a larger number of iterations, integrate over the expectation operators.

\begin{align}
    L^{i,j}_{ee} =& \left(1 - \mu^i_t - \beta R_t \left[ \frac{exp(\psi^{i, j}_{t+1})}{exp(\psi^{i, j}_t)} \left( \frac{\lambda^i_t}{\lambda^{i, j}_{t+1}} \right)^\sigma \frac{1}{\Pi^j_{t+1}} \right]\right) \\ 
    L^j_{nkpc} =& \left( \phi \left( \frac{\Pi_t}{\Pi} - 1 \right) - (1 - \epsilon) - \epsilon MC_t - \beta \phi \mathbb{E}_t \left[ \frac{\Pi^j_{t+1}}{R_t} \left( \frac{\Pi^j_{t+1}}{\Pi} - 1 \right) \frac{\Pi^j_{t+1}}{\Pi} \frac{Y^j_{t+1}}{Y_t} \right] \right) \\
    L^i_{ls} =& \left( \left(c^i_t - hC_{t-1})\right)^{-\sigma} - \left( \frac{\chi \left( h^i_t \right)^\eta}{s^i_t W_t} \right) \right) \label{labour_supply_loss} \\
    L =& \frac{1}{L} \sum_{i=1}^L \left[ L^{i, 1}_{ee} \cdot L^{i, 2}_{ee} + \left( L^i_{ls} \right)^2\right] + L^1_{nkpc} \cdot L^2_{nkpc} \label{eqn:hank_loss}
\end{align}

When the outputs are constrained, in the \textit{hard-constraint} approach, these are all the loss components that are necessary. On the other hard, when the outputs are not constrained, a further $3$ loss components are required, to ensure that deviations from the resource and budget constraints are sufficiently penalised. These penalties are given in equations (\ref{eqn:bc_loss}), (\ref{eqn:oc_loss}), and (\ref{eqn:rc_loss}).

\begin{align}
    L^i_{bc} =& \Psi_{fb}\left( b^i_t - \underbar{B}, \mu^i_t \right) \label{eqn:bc_loss} \\
    L_{oc} =& Y - \frac{1}{L} \sum_{i=1}^L c^i_t  \label{eqn:oc_loss} \\
    L_{rc} =& \frac{1}{L} \sum_{i=1}^L b^i_t \label{eqn:rc_loss}
\end{align}

\subsection{Learning Algorithm}

The \textit{ADAM} learning algorithm \parencite{kingma2014adam} will be used to learn the neural-network parameters. Batches of size $mb$ each containing states and randomly drawn structural parameters are fed into the neural-network, and proposed policies are calculated. From this, the loss and gradient of the loss are evaluated and the update step is performed. After this new random structural parameters are drawn, and the model is simulated forward up 20 times according to the state-transition rule of the model, and the current approximated policy. The resulting states and structural parameters are then used as the inputs in the next iteration. This procedure is repeated until a desired number of iterations are reached, or until the loss reaches a sufficiently small value.

When constraints are soft, if any of the constraint related losses go above a certain thresh-hold the states are reset to initial values, and the number of forward simulations between each update is decreased. Conversely, when the constraint related losses go below some low thresh-hold the number of forward iterations is increased, up to a maximum of 20. This procedure ensures that the approximation is stable, especially for initial values, where it can easily diverge. These steps are summarised in Algorithm \ref{alg:fitting} in the appendix.

\subsection{Constraint Satisfaction} \label{sec:constraint_satisfaction}

This section will explain how to apply the transformations in Section \ref{methods} in order to assure that all of the constraints of the model are satisfied. This model has in effect six constraints, four idiosyncratic constraints: (\ref{budget_constraint}) (\ref{borrowing_constraint}), (\ref{consumption_constraint}) and (\ref{labour_constraint}), and two aggregate or market clearing constraints: (\ref{resource_constraint}) and (\ref{output_constraint}). First ensure output for $h^i_t$ is positive by applying a softplus activation function. With these, the predicted prices $W_t$, $\Pi_t$, and the states of the model we can calculate $N_t$, $Y_t$, $MC_t$ and, $Div^i_t$, and thus the total wealth of every agent before consumption $\omega^i_t$. Then (\ref{budget_constraint}), (\ref{output_constraint}), (\ref{resource_constraint}) imply that $\frac{1}{L} \sum_{i=1}^L \omega^i_t = \frac{1}{L} \sum_{i=1}^L c^i_t + (0) = Y_t$. Therefore, the remaining constraints can all be satisfied by choosing $c^i_t$ for each agent such that:

\begin{align}
    \overbrace{0}^{a_i} < c^i_t &\leq \overbrace{\omega^i_t - \underline{B}}^{b_i} \label{cons1} \\
    \sum_i^{L} c^i_t &= \underbrace{\sum_i^{L} \omega^i_t}_{C} \label{cons2}
\end{align}

This is a constraint is of the general type outlined in Section \ref{methods} and therefore the algorithm outlined there can be used to re-scale $\tilde{c}^i_t$ into valid choices. In particular we use the variation where the constraint on $b^i$ can bind, so that a mass of agents can hit their budget constraint exactly. Given the resulting $b^i_t$ we can evaluate if the budget constraint binds $b^i_t = \underbar{B}$ for every agent, and then assign their Lagrange multiplier accordingly: $\mu^i_t = \mathbbm{1}\{ b^i_t = \underbar{B} \} \tilde{\mu}^i_t$.

I also consider two intermediate cases. The first of which is the case in which the idiosyncratic constraints (in this case, the budget constraint (\ref{budget_constraint}) and borrowing constraint (\ref{borrowing_constraint})) are satisfied by construction, but no re-scaling is done, so market clearing constraints (Equations (\ref{output_constraint}) and (\ref{resource_constraint})) are represented only by the penalties (\ref{eqn:oc_loss}) and (\ref{eqn:rc_loss}). Here too, the exact implementation details matter. One approach to doing this would be to take a consumption choice $\tilde{c}^i_t$ in $(0, 1)$ by using a sigmoid activation function, and then multiplying this by the financial capacity of each agent in order to get a valid consumption choice relative to the budget constraint, as shown in (\ref{eqn:idio_only_sigmoid}).

\begin{align}
    &c^i_t = (\omega^i_t - \underbar{B}) \cdot \tilde{c}^i_t \nonumber \\ 
    &\tilde{c}^i_t = \frac{1}{1+e^{-Z_t}} \in (0, 1) \label{eqn:idio_only_sigmoid}
\end{align}

However, constraining consumption in this initial way would present a number of problems. Firstly, since the consumption choice is strictly bound from above, none of the agents could ever actually hit their budget constraint. This could be alleviated by using a \textit{hard sigmoid} function, but this would not get around the second problem which is that the gradient of such consumption functions is close to or exactly zero when the agent is at the budget constraint. As a result, the gradient descent optimiser is essentially unable to learn about optimal choices close to or at the constraints. This is a common problem in applied machine-learning known as the \textit{vanishing gradient} problem. Instead, in order to avoid this, a strictly positive consumption choice is taken for each agent via a softplus activation function, and it is simply clipped at the financial capacity of each agent, as shown in (\ref{eqn:idio_only_clip}). This allows for consumption choices at the budget constraint to be common without inducing a vanishing gradient.

\begin{align}
    &c^i_t = \min\left\{ \tilde{c}^i_t, \omega^i_t - \underbar{B} \right\} \label{eqn:idio_only_clip} \\ 
    &\tilde{c}^i_t = log(1 + exp(Z_t)) \in (0, \infty) 
\end{align}

The other intermediate case is one in which the aggregate constraints  (\ref{resource_constraint}) and (\ref{output_constraint}) are satisfied by construction, and the borrowing constraint (\ref{borrowing_constraint}) is implemented via the FB penalty. This is achieved by a re-scaling in which again each agent picks a strictly positive consumption level $\tilde{c}^i_t$, and their sum is re-scaled such that:

\begin{align}
    c^i_t =& \frac{1}{L} \sum_{i=1}^L \omega^i_t \cdot \frac{\tilde{c}^i_t}{\frac{1}{L} \sum_{i=1}^L \tilde{c}^i_t} \\
    \implies \frac{1}{L} \sum_{i=1}^L c^i_t =&  \frac{1}{L} \sum_{i=1}^L \omega^i_t = Y_t \nonumber
\end{align}

In the soft-constraint approach, there is no need to calculate wealth before calculating consumption, so the labour supply FOC (\ref{labour_supply_equation}) can be substituted in directly, such that the labour supply is exactly optimal for any given consumption choice.

\subsection{Calibration}

\begin{table}
    \centering
    \begin{tabular}{|c|c|c|c|}
        \hline
        Parameter & Baseline Value & Min & Max \\
        \hline
        $\beta$ & 0.9975 & 0.9975 & 0.9975 \\
        \hline
        $\sigma$ & 1 & 1 & 1 \\
        \hline
        $\eta$ & 1 & 1 & 1 \\
        \hline
        $\epsilon$ & 11 & 11 & 11 \\
        \hline
        $\chi$ & 0.91 & 0.91 & 0.91 \\
        \hline
        $h$ & 0 & 0 & 0 \\
        \hline
        $\phi$ & 1000 & 700 & 1300 \\
        \hline
        $\theta_\Pi$ & 2 & 1.5 & 2.5 \\
        \hline
        $\theta_Y$ & 0.25 & 0.05 & 0.5 \\
        \hline
        $\Pi$ & 1.005 & 1.005 & 1.005 \\
        \hline
        $Y$ & 1 & 1 & 1 \\
        \hline
        $\underbar{B}$ & -0.05 & -0.5 & -0.01 \\
        \hline
        $\rho_\Psi$ & 0.7 & 0.5 &  0.9 \\
        \hline
        $\rho_s$ & 0.8 & 0.7 & 0.9 \\
        \hline
        $\rho_A$ & 0.8 & 0.7 & 0.9 \\
        \hline
        $\rho_r$ & 0.25 & 0.1 & 0.5 \\
        \hline
        $\sigma_\Psi$ & 0.03 & 0.01 & 0.05 \\
        \hline
        $\sigma_s$ & 0.05 & 0.01 & 0.08 \\
        \hline
        $\sigma_A$ & 0.008 & 0.003 & 0.012 \\
        \hline
        $\sigma_{mp}$ & 0.005 & 0.001 & 0.008 \\
        \hline
    \end{tabular}
    \caption{Calibration of parameters used to approximate the model in Section \ref{model}, along with upper and lower bounds between which parameters are uniformly drawn during training.}
    \label{tab:calibration}
\end{table}

Table \ref{tab:calibration} displays the parameters used for approximating the model. Many of these are taken from \citeauthor{kase2022estimating} (\citeyear{kase2022estimating}), with the exception of $h$, which was set to zero, $\underbar{B}$ and $\sigma_s$, which were increased to ensure that many agents hit their borrowing constraint in equilibrium, and $\rho_A$ and $\sigma_A$, which were added as I model the TFP process as deviations from steady-state rather than trend growth. As in that paper, the economic parameters are added as inputs to the neural-network, and during training they are occasionally uniformly re-sampled between the given bounds. When the lower and upper bounds are the same, then that parameter is unchanged, essentially it is calibrated. Although this paper does not deal directly with the estimation of structural parameters, they are still included in the policy approximation in order to keep this possibility open in extensions and also because I find qualitatively that varying the structural parameters may actually improve rate of convergence of the fitting algorithm. This could be because exposing the neural-network to different combinations of structural parameters may facilitate more general learning about the problem space. Furthermore, approximating the model over a range of parameters in this way makes it possible to query the model in different ways over ranges of parameters without refitting the model, which saves a considerable amount of time.

\subsection{Calculation of Impulse Responses}

Figures \ref{fig:soft_hard_irf} and \ref{fig:agg_idio_irf} display generalised IRFs \parencite{koop1996impulse} generated by first taking a sample of states from the ergodic distribution and then for each of these states a large number of simulations are performed, in each of which the same shock (i.e. $\epsilon^A_t = 2$) is introduced at $t=0$, and the state-transition is simulated forward in the standard way, where other shocks are not turned off, but are instead drawn from their respective distributions, as usual. The generalised IRF is then the average response of the outcome variables over this large number of draws of shocks. The generalised IRFs in these figures were generated using 4000 draws for each of 256 states, resulting in over 1 million simulated impulse responses in total. Despite this, there is still a noticeable amount of noise in the impulse responses. This could be attenuated by simulating even more impulse responses, however, the approximation error is likely to remain meaningful for any feasible number of simulations, since there are a large number of shocks to integrate over. 

Using generalised IRFs is necessary, not only because the solution is non-linear, but also because with the neural-network solution method we do not explicitly solve for any deterministic steady-state. Instead, we solve the model globally for idiosyncratic and aggregate shocks simultaneously. Indeed, since we allow for the possibility that the distribution of idiosyncratic shocks can cause aggregate fluctuations in our solution, there is no deterministic steady-state as such unless we turn off idiosyncratic shocks, which are the only source of heterogeneity in the model considered here.\footnote{This is why it makes sense to eschew \citeauthor{greenwood1988investment} (\citeyear{greenwood1988investment}) preferences.} Instead, this method finds the \textit{ergodic distribution} of states. The generalised IRF integrates over the distribution of future shocks and current states to display an approximate average response to a particular shock. The advantage of this is that we can easily see how the response varies over different initial states of the economy. Since the model is not linearized and is instead solved in levels, for ease of comparison the IRFs are transformed into deviations by subtracting the mean and dividing by the standard deviation of each policy, which were calculated from a simulated time-series.

\section{Results} \label{results}

\subsection{Hard and soft-constraints} \label{output_constraints}

\begin{table}
    \centering
    \begin{tabular}{|l|l|c|c|}
        \hline
        & & \bfseries All Hard & \bfseries All Soft  \\
        \bfseries Type & & \bfseries Constraints & \bfseries  Constraints  \\ \hline
        & Total Loss & 3.60e-04 & 1.13e-02 \\ \hline
        & Euler Equation Loss & 2.94e-04 & 5.64e-03 \\ 
        Optimality & Phillips Curve Loss & 1.17e-07 & 5.14e-03  \\
        & Labour Supply Loss & 6.64e-05 & 1.10e-32  \\ \hline
        & KKT Loss & 3.25e-35 & 2.74e-04  \\ 
        Constraints & Output Constraint Loss & 3.96e-32 & 1.03e-04  \\
        & Net Supply of Bonds & 1.36e-32 & 1.56e-04 \\ \hline
    \end{tabular}
    \caption{Average total loss and its constituent components evaluated over the last 50 iterations of the fitting procedure for the models fit with all constraints hard and all constraints soft (penalty-based). The soft-constraint version was fit with weights of $1e2$ on each of the three constraint-related loss components.}
    \label{tab:soft_hard_losses} 
\end{table}    

\begin{figure}
    \centering
    \captionsetup{width=0.8\textwidth}
	\includegraphics[width=0.8\textwidth]{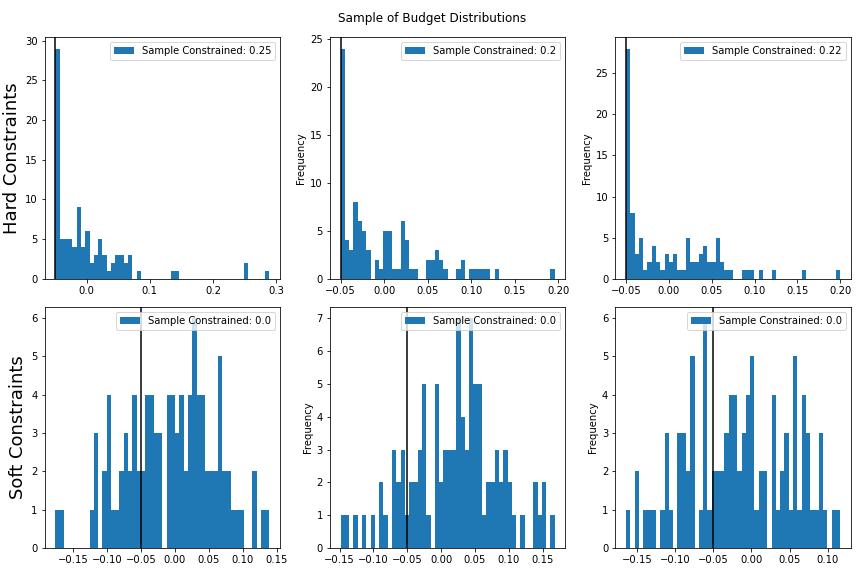}
	\caption{Bond holding distributions for 3 randomly sampled states (across columns) for the hard-constraint method (top row) and soft-constraint method (bottom row).}
	\label{soft_vs_hard_budgets}
\end{figure}

Table \ref{soft_vs_hard_budgets} compares the losses obtained by the hard and soft-constraint solution methods. In this average over a range of different calibrations the hard-constraint method handily outperforms the soft-constraint one. Note that losses that are less than $1e-30$ can be attributed to floating-point imprecision; these are the conditions that are fulfilled exactly for each solution method. For the hard-constraint version this is all three constraints, and for the soft-constraint version this is the labour supply loss, which can be fulfilled by construction using the labour-consumption FOC when other constraints are not taken into account. The hard-constraint implementation achieves an overall loss almost two orders of magnitude smaller than the soft-constraint approach. However, in the soft-constraint case, the larger overall loss does not come from the constraint penalties themselves, which are all at least an order of magnitude smaller than the other loss components. Rather, the presence of relatively small, but still meaningful errors in the constraint conditions makes it more difficult for this method to satisfy the FoCs of the model. In particular, the soft-constraint has its worst relative performance on the Phillips curve loss, which is an equation that involves only aggregate variables. As discussed, this is could be because the states that are generated as inputs are further from the true ergodic equilibrium, or because the optimiser is trading off these optimality conditions with the essentially arbitrary penalties that it incurs.

\begin{figure}
    \centering
    \captionsetup{width=0.8\textwidth}
	\includegraphics[width=0.8\textwidth]{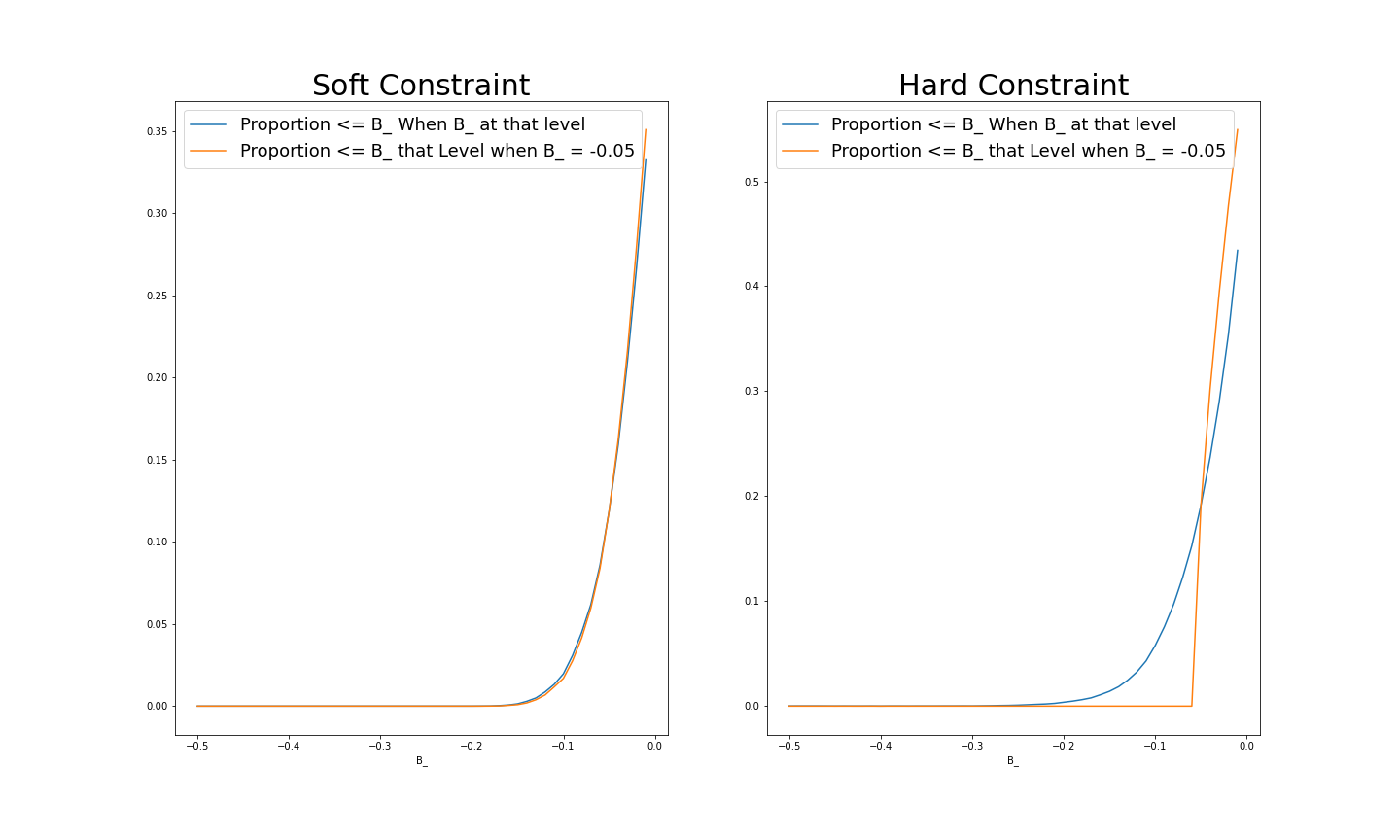}
	\caption{The average proportion of agents whose end of period savings are at or below a given value of the budget constraint when that value is the value of $\underbar{B}$ (blue line) and the average proportion of agents whose end of period savings fall below a given value when the budget constraint is set to the default value of $-0.05$ (yellow line) for the soft-constraint model (left) and the hard-constraint model (right).}
	\label{soft_vs_hard_relative}
\end{figure}

Not only does the hard-constraint version provide quantitatively much better results, but this solution is also qualitatively better in a number of critical ways. For example, it is reasonable to expect that the proportion of constrained agents is increasing in the borrowing constraint. The hard-constraint version of the model captures this correctly, while the soft-constraint struggles. This can be seen in Figure \ref{soft_vs_hard_relative}. In this figure the yellow line depicts the average proportion of agents from a sample of draws from the ergodic set who fall at or below a given level of wealth when the budget constraint is set to its default value of $-0.05$, in other words, the cumulative distribution of wealth when $\underbar{B} = -0.05$. The blue line depicts the average proportion of agents who are at the constraint $\underbar{B}$ at that value as the constraint varies across values on the x-axis. 

\begin{figure}
    \centering
    \captionsetup{width=0.8\textwidth}
	\includegraphics[width=0.8\textwidth]{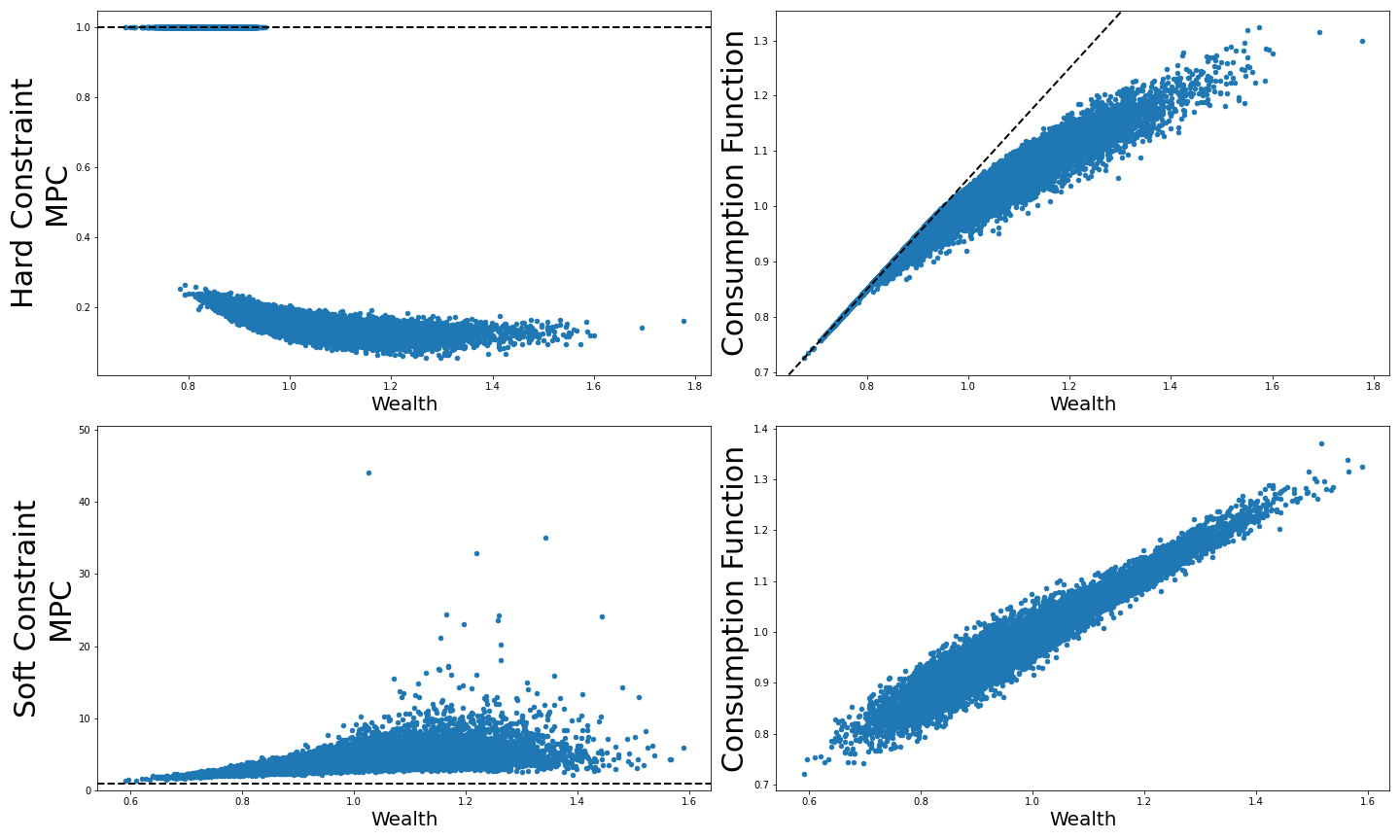}
	\caption{Immediate MPCs (right column) and consumption functions (left column) for the hard-constraint method (top row) and soft-constraint method (bottom row) across all agents in a sample of states drawn from the ergodic distribution.}
	\label{soft_vs_hard_mpc}
\end{figure}

For the soft-constraint version the two lines are almost identical. This tells us that this model fit with soft-constraints has learned relatively little about how agents' behaviour should change around the constraint, as the shape of the distribution of wealth is essentially invariant to the value of $\underbar{B}$. Therefore, although the soft-constraint model does predict less agents are constrained when the constraint is weaker, this is simply a by-product of sliding the value of $\underbar{B}$ along the same wealth distribution, rather than deeper learning about the implications of the budget constraint. On the other hand, the hard-constraint version does not allow any agents to have wealth below $-0.05$ when $\underbar{B} = -0.05$ (yellow line), while it does predict a smoothly decreasing proportion of constrained agents as the budget constraint is weakened (blue line). This figure also illustrates another important outcome: imposing hard-constraints does not force the constraint to bind. It is clearly possible for exactly zero of the agents to be at the budget constraint, and indeed this is the case when the budget constraint is weak. This difference in the handling of the budget constraint is important because the high MPC of agents near to the budget constraint is exactly what causes the implications of the HANK model to deviate from those of a RANK model \parencite{kaplan2018monetary}. MPCs generated by each approach are displayed slightly later in Figure \ref{soft_vs_hard_mpc}.

\begin{figure}
    \centering
    \captionsetup{width=0.8\textwidth}
	\includegraphics[width=0.8\textwidth]{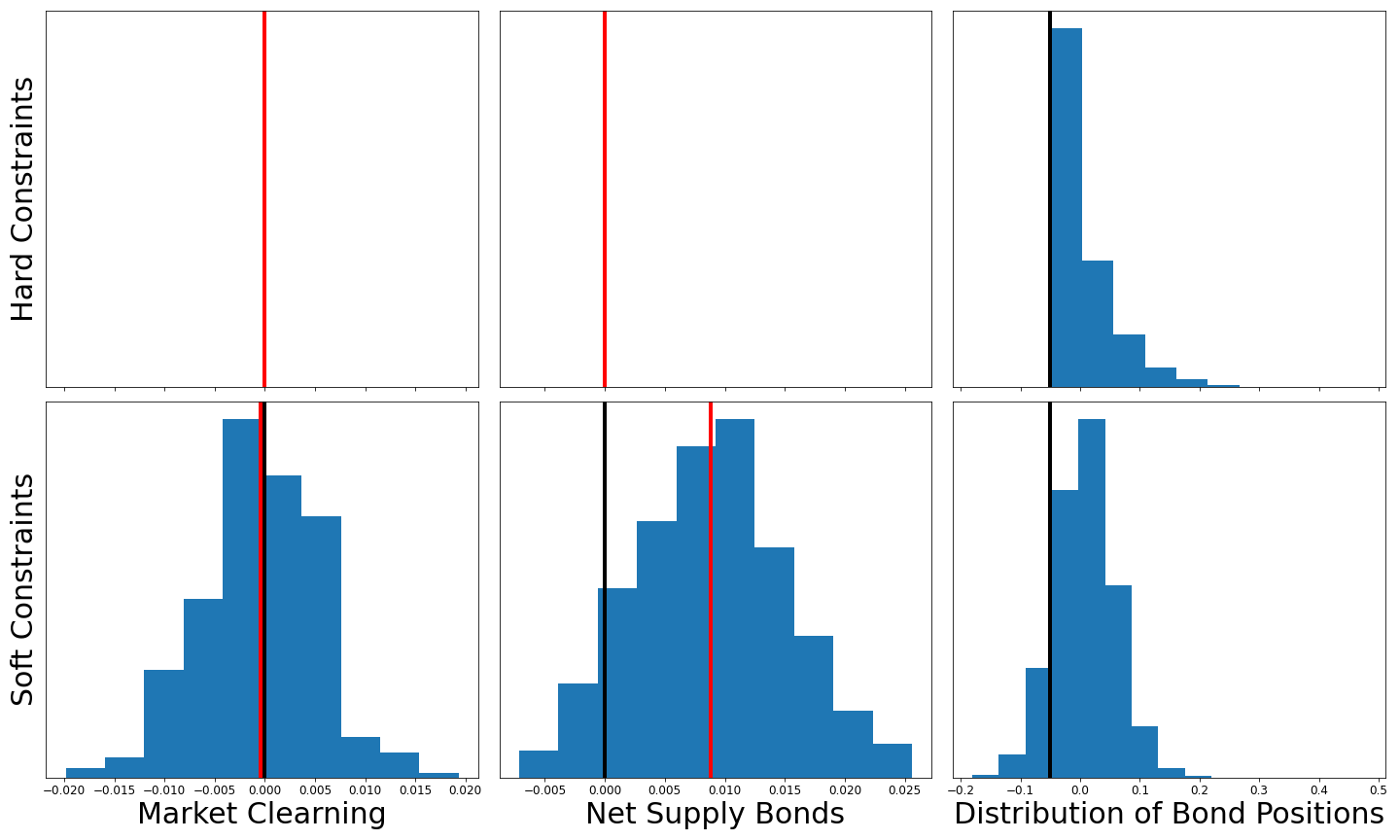}
	\caption{Cross-Sectional distribution of output net of consumption (leftmost column), net supply of bonds (middle column), and individual bond positions (rightmost column) for the hard-constraint implementation (top row) and soft-constraint implementation (bottom row). Histogram spans observations from a batch of 256 simulated draws from the ergodic set. Red lines indicate means over the batch.}
	\label{soft_vs_hard_constraints}
\end{figure}

Figure \ref{soft_vs_hard_constraints} visually demonstrates the distribution of errors related to the three constraints in the model combined from a batch of states drawn from the approximated ergodic distribution for the two different methods. The top row displays the results for the hard-constraint method, and we can see that the market clearing error and net bond supply are imperceptible when put on the same scale as the error from the soft-constraint model. The rightmost column shows the distribution of idiosyncratic bond positions for all agents across the drawn states where the borrowing constraint is indicated by the vertical black line. There are striking differences in the shape of the wealth distribution generated by both methods. For the hard-constraint method, around $20\%$ of the observations lie exactly on the borrowing constraint, and the distribution is shaped as a decay to the right. For the soft-constraint method, many observations lie to the left of the borrowing constraint, and the distribution has a significantly more symmetrical shape. Furthermore, we see that the soft-constraint version seems have a notable positive bias in the net supply of bonds. The optimiser could have chosen to do this in order to trade off the resulting penalty with a reduced penalty from the budget constraint, as the positive net supply of bonds makes it less likely that agents will violate the budget constraint, all else equal. These notable issues are despite the fact that the errors related to the model constraints are of the order $1e-4$ for the soft-constraint method, which is small relative to the errors in the first-order conditions, even for the hard-constraint model. So while this error is quantitatively small, it is still clearly economically relevant. The difference in the distribution of bond holdings can be seen even more clearly when comparing the distribution of bond positions for all agents in a sample of individual states for the hard and soft-constraint methods, as shown in Figure \ref{soft_vs_hard_budgets}.

Figure \ref{soft_vs_hard_mpc} shows a stark difference in the immediate MPCs generated by the hard and soft-constraint approaches.\footnote{For the hard-constraint approach these MPCs can be calculated directly from the re-scaling function, which is a differentiable function that takes wealth as an input and outputs consumption. On the other hand, for the idiosyncratic and soft-constraint versions, the consumption policy is only a function of states. Therefore, immediate MPC is calculated as $\frac{dc^i_t}{d\omega^i_t} = \frac{dc^i_t}{db^i_{t-1}}/\frac{d\omega^i_t}{db^i_{t-1}}$. In order to maintain consistency, MPCs for both approaches are calculated in this latter way.} The hard-constraint method (first row) faithfully reproduces consumption behaviour which is qualitatively consistent with the mechanisms of the HANK model. The consumption function is kinked at the bottom as a result of the borrowing constraint\footnote{This is a standard feature of heterogeneous agent models with incomplete markets, such as \citeauthor{krusell1998income} (\citeyear{krusell1998income}), for example.} and there is considerable heterogeneity in immediate MPCs: constrained agents, that is to say agents who ended the period with bond positions at the budget constraint, and who therefore mostly have lower total wealth, have MPCs of 1, while unconstrained agents have MPCs decreasing in wealth at a much lower mean of around 0.2. 

This discontinuous change in behaviour is exactly the kind of behaviour that the soft-constraint approach struggles to model correctly. Indeed, the soft-constraint (bottom row) version does not reproduce any of the qualitative features that we expect a solution to display. The consumption function appears to be nearly linear, and many of the MPCs are much greater than 1, and appear to be increasing in wealth, which is inconsistent with optimising behaviour in this setup. This comes as a direct result of the failure of the model to strongly enforce the budget constraint. 

\begin{figure}
    \centering
    \captionsetup{width=0.8\textwidth}
    \includegraphics[width=0.8\textwidth]{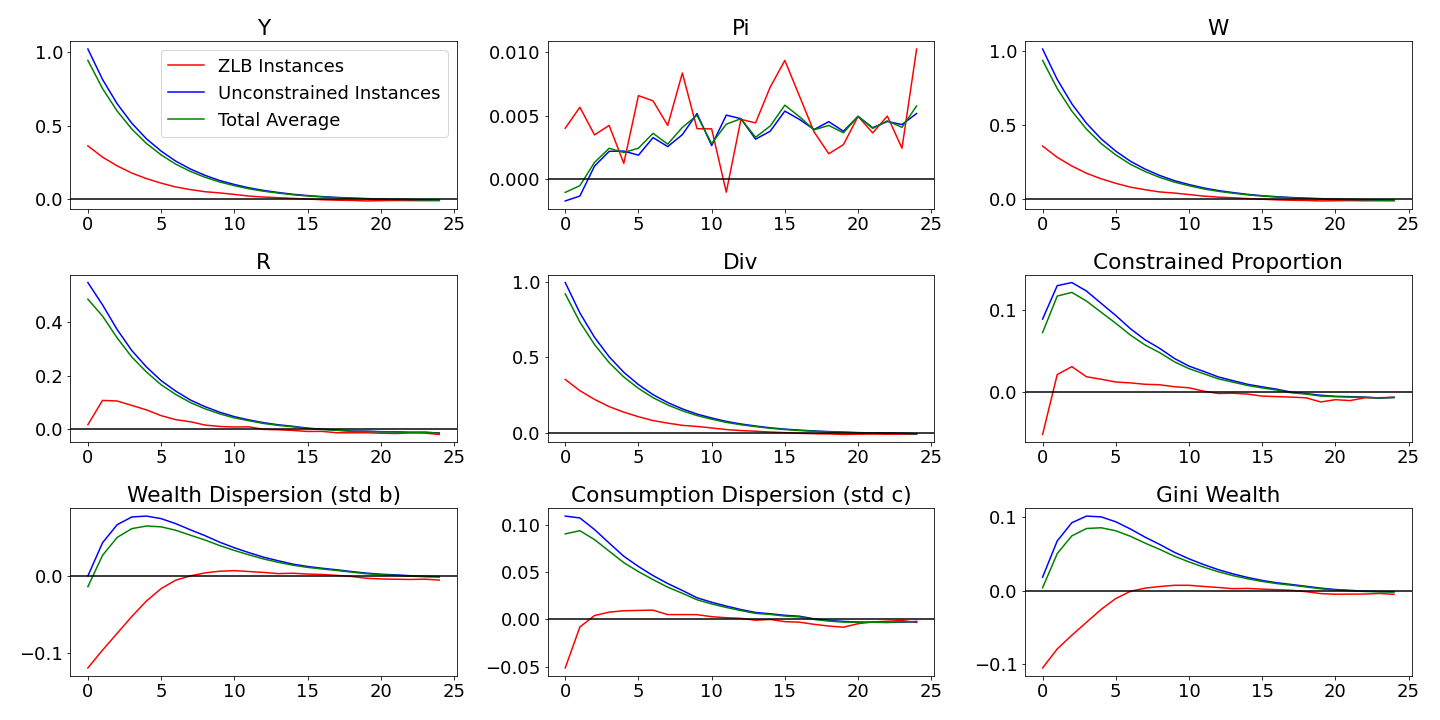}
    \includegraphics[width=0.8\linewidth]{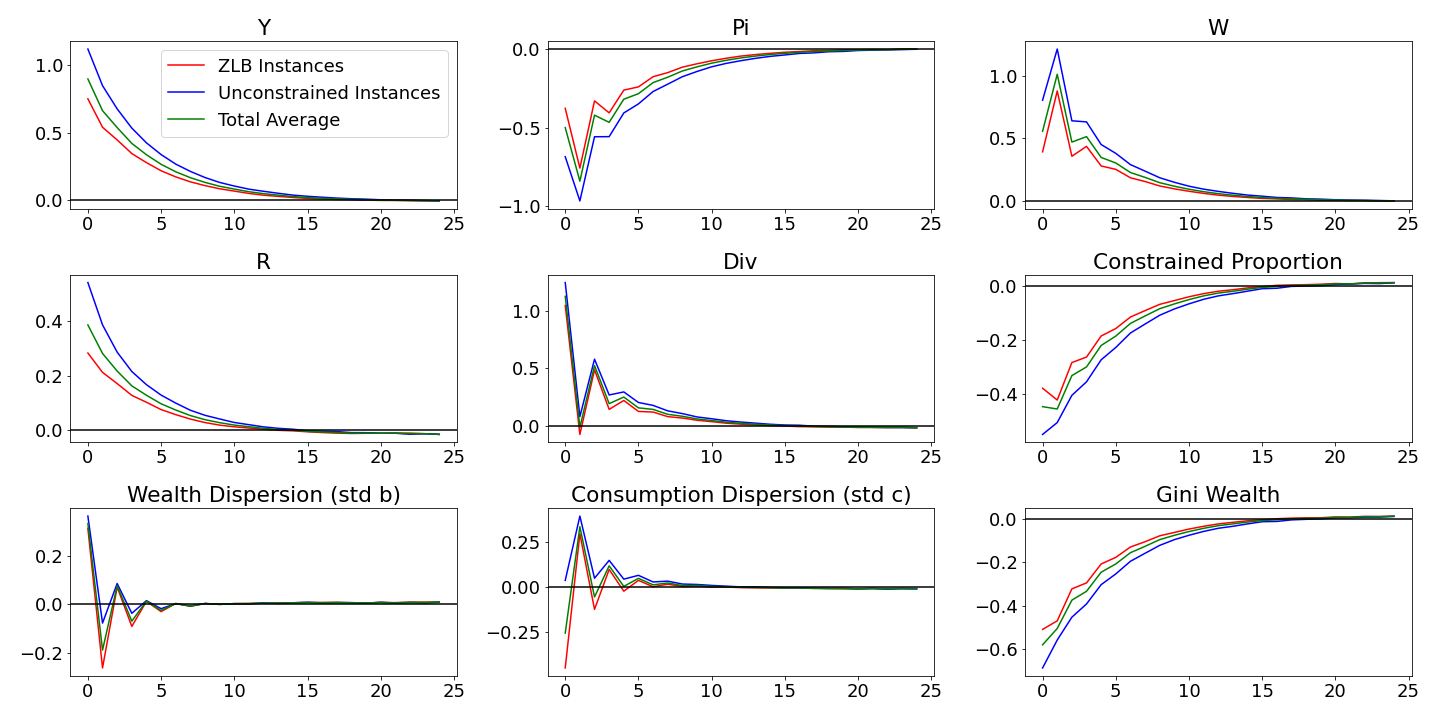}
	\caption{Impulse Responses to a 2 standard deviation expansionary TFP shock ($\epsilon^A_t$) generated from the hard-constraint (top three rows) and soft-constraint (bottom three rows) neural-networks. Note that for the soft-constraint version the constrained-proportion is the proportion that fall outside the budget constraint.}
	\label{fig:soft_hard_irf}
\end{figure}

Figure \ref{fig:soft_hard_irf} shows the generalised impulse responses to a two standard deviation expansionary TFP shock under each of the two main constraint regimes. The responses in red are for ZLB episodes (i.e. when $R_{t-1} = 0$), while the blue are for all other states, and green shows the total average over all states in the batch. The response of wealth dispersion for the hard-constraint model, for example, demonstrates one of the major benefits of the neural-network solution method. The response has a different direction and shape in the ZLB states compared to the unconstrained states. This shows how the solution is able to capture responses which are non-linear in the aggregate. 

When comparing these, the hard-constraint version (top) should be regarded as closer to the ground-truth, given its much lower loss, and thus, approximation error.\footnote{However, one response that seems problematic is that of inflation. This erratic behaviour is primarily an artefact of the calculation of generalised impulse responses. In the solution inflation moves very little (with a standard deviation of the order $1e-6$ because the monetary authority has the ability to fully stabilise in this setup. As a result of this low signal-to-noise ratio when the response is normalised it becomes pure noise.} The soft-constraint version struggles to match the impulse responses of the hard-constraint solution in both the aggregate and idiosyncratic variables. This is in part because it predicts that the zero lower bound is encountered far too frequently. This can be seen clearly in the distribution of prices and aggregate variables shown in Appendix \ref{appendix:agg_policies}. In particular, it has a hard time capturing aspects of the cross-sectional distribution. The responses of the constrained proportion, standard deviation of $b^i_t$, standard deviation of $c^i_t$ and the gini-coefficient are all substantially different than the hard-constraint model. Furthermore, the soft-constraint model seems to predict essentially no impact of the shock on the dispersion of wealth or consumption, which is implausible.

Of course, it is still possible that the results for the soft-constraint method could be made better by using a different neural-network architecture, using different step-sizes and weights on loss components, training longer, or picking the initial conditions differently. To more closely consider one of these possibilities, Appendix \ref{appendix:soft_weights} provides results obtained by training the soft-constraint version of the model with varying weights on the constraint-related penalties. The results vary relatively little over this dimension when models are allowed to train for a large number iterations, and all versions still perform substantially worse than hard-constraints. Notwithstanding this, it is quite possible that there is some combination of weights that performs even better. The main problem is that these neural-network fits do require a substantial amount of time to run, so it is undesirable to have hyper-parameters such as the weights on various loss components that need to be fine-tuned in order to obtain a reliable approximation. While implementing hard-constraints can also be complicated and time-consuming for the user, this paper already provides a solution for a large class of problems. 

In order to be as fair as possible as many factors were kept the same as possible, modifications were made that allowed the soft-constraint version to produce the best results possible (outlined in Section \ref{methods}), and the soft-constraint version was even allowed to train significantly longer (see Figure \ref{tab:settings} in the appendix for details). The purpose of these results is not to argue that the soft-constraint method does not work at all, but rather to demonstrate that the proposed hard-constraint method is more accurate and converges faster, and is therefore a better alternative. 

\subsection{Intermediate Cases}\label{sec:intermediate_cases}

In order to demonstrate the effectiveness of the hard-constraint framework, this section will also compare the approximated policy functions generated by two alternative, intermediate methods. The first, as outlined in Section \ref{sec:constraint_satisfaction} is an intermediate approach, in which re-scaling ensures that market-clearing is satisfied by construction, and the budget constraint is implemented by the Fischer-Burmeister penalty. This is henceforth referred to as the aggregate-constraint version. The second, as in \citeauthor{maliar2020deep} (\citeyear{maliar2020deep}) is the converse intermediate approach, where the the borrowing constraint is satisfied by construction by clipping consumption choices at their feasible maximums, and market-clearing is enforced through a penalty. This is referred to as the idiosyncratic-constraint version.

\begin{table}
    \centering
    \begin{tabular}{|l|l|c|c|c|c|}
        \hline
        & & \bfseries All Hard & \bfseries All Soft & \bfseries Hard Agg & \bfseries Hard Idio  \\
        \bfseries Type & & \bfseries Constraints & \bfseries  Constraints & \bfseries Constraints & \bfseries Constraints  \\ \hline
        & Total Loss & 3.60e-04 & 1.13e-02 & 2.09e-03 & 7.07e-02 \\ \hline
        & Euler Equation Loss & 2.94e-04 & 5.64e-03 & 6.51e-04 & 4.89e-03 \\ 
        Optimality & Phillips Curve Loss & 1.17e-07 & 5.14e-03 & 1.09e-03 & 4.06e-02  \\ 
        & Labour Supply Loss & 6.64e-05 & 1.10e-32 & 3.52e-04 & 2.50e-02  \\ \hline
        & KKT Loss & 3.25e-35 & 2.74e-04 & 1.87e-06 & 1.33e-34 \\ 
        Constraints & Output Constraint Loss & 3.96e-32 & 1.03e-04 & 5.25e-32 & 6.62e-05 \\ 
        & Net Supply of Bonds & 1.36e-32 & 1.56e-04 & 1.38e-32 & 1.25e-04 \\ \hline
    \end{tabular} 
    \caption{Average total loss and its constituent components evaluated over the last 50 iterations of the fitting procedure for the models fit with only aggregate constraints hard and only idiosyncratic constraints hard. Previous results are repeated for ease of comparison.}
    \label{tab:intermediate_losses}  
\end{table}

Table \ref{tab:intermediate_losses} shows the overall loss, and components thereof generated by the intermediate approaches. Given that the most salient issues with the soft-constraint model seemed to be related to the budget constraint, one might expect to see that the idiosyncratic constraint model was the best intermediate approach. However, interestingly, we see that the aggregate-constraint and the soft-constraint models actually out-perform the idiosyncratic-constraint model. Despite these small losses with regard to constraints, the aggregate policies generated by the idiosyncratic-constraint method are highly implausible, for example, we can see in Figure \ref{fig:agg_dists_agg_idio} in the appendix that this model produces an average output of only $0.94$ (the true equilibrium value, which is the centre of the distribution generated by the hard-constraint and soft-constraint models is $1$), and therefore, strikes the ZLB far too often.

\begin{figure}
    \centering
    \captionsetup{width=0.8\textwidth}
	\includegraphics[width=0.8\textwidth]{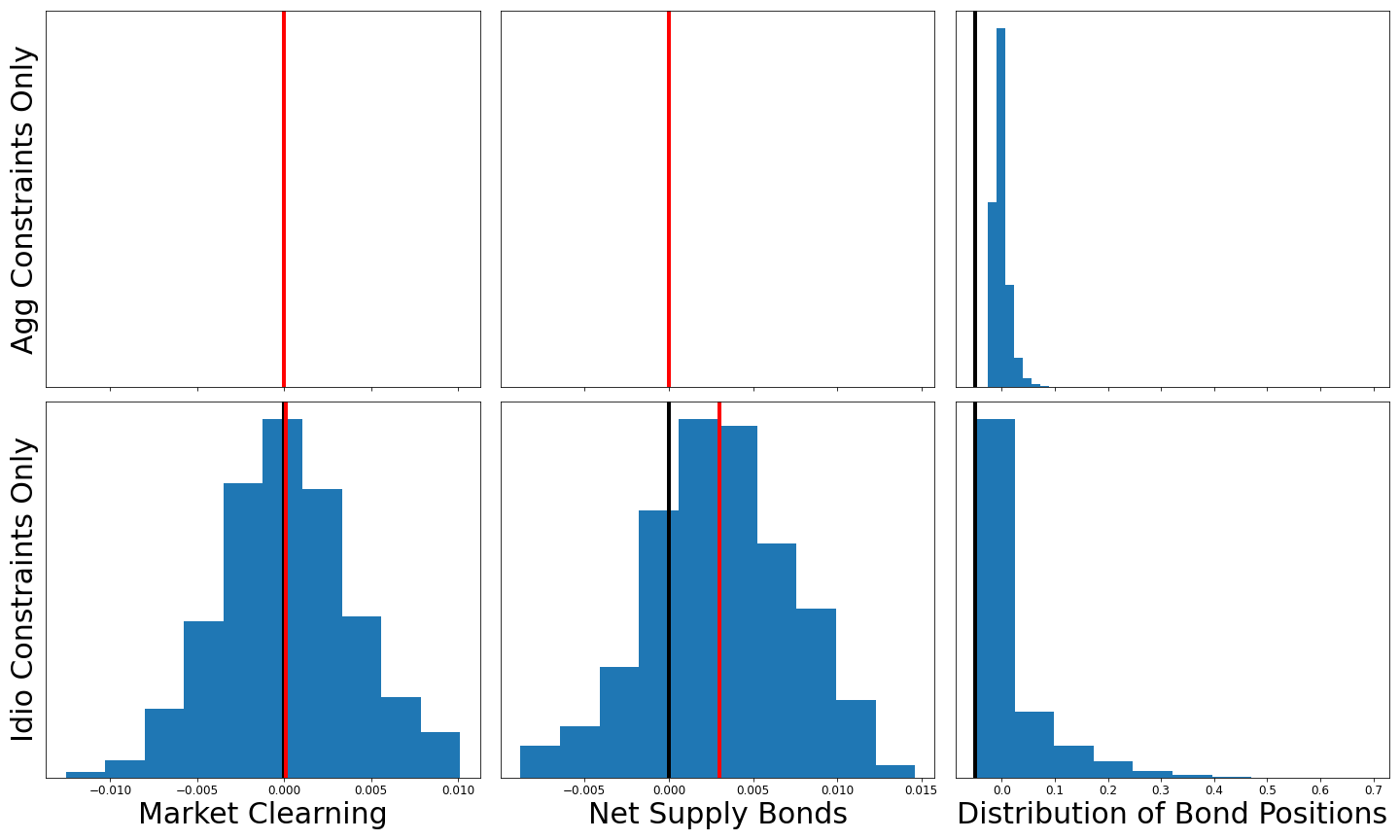}
	\caption{Cross-Sectional distribution of output net of consumption (leftmost column), net supply of bonds (middle column), and individual bond positions (rightmost column) for the implementation where only the aggregate constraints are hard (top row), and implementation where only the idiosyncratic constraint is hard (bottom row).}	\label{fig:agg_vs_idio_constraints}
\end{figure}

The aggregate-constraint solution, on the other hand, performed much better, and did manage to out-perform the soft-constraint version, although its overall loss was still an order of magnitude larger than for the hard-constraint method. In this case it is interesting to note that the KKT loss was also significantly lower than in the soft-constraint version. This is likely because enforcing aggregate constraints creates a virtuous feedback cycle in which improvements to the approximation of aggregate variables and prices subsequently improves in the approximation of idiosyncratic variables, since the states being fed into the neural-network are closer to the true equilibrium. However, the aggregate-constraint model, even with a KKT Loss of the order $1e-6$ still does not capture aspects of the cross-sectional distribution as well as the hard-constraint model. This can be seen in the budget distributions in Figure \ref{fig:agg_vs_idio_constraints}.

As expected the aggregate-constraint version produces imperceptible errors in the aggregate constraints, while the idiosyncratic-constraint version succeeds in generating a budget distribution where the budget constraint binds frequently but is never violated. However, in contrast to the soft-constraint version, where the budget constraint was often violated, the aggregate-constraint version produces outputs that never touch budget constraint. This model has the opposite problem, which is that the budget constraint never binds, although we know that this is an equilibrium outcome. This reflects a tendency that arises when using the Fischer-Burmeister penalty: in cases where the penalty for breaking the constraints is large relative to other loss components the solution may produce outputs strictly inside the constraint. This is related to the tendency of the FB function to under-punish interior solutions, as highlighted by \citeauthor{chen2000penalized} (\citeyear{chen2000penalized}.). On the other hand, in cases where the penalty from breaking constraints is relatively small the model may generate outputs that ignore the constraints entirely. 

\begin{figure}
    \centering
    \captionsetup{width=0.8\textwidth}
    \includegraphics[width=0.8\textwidth]{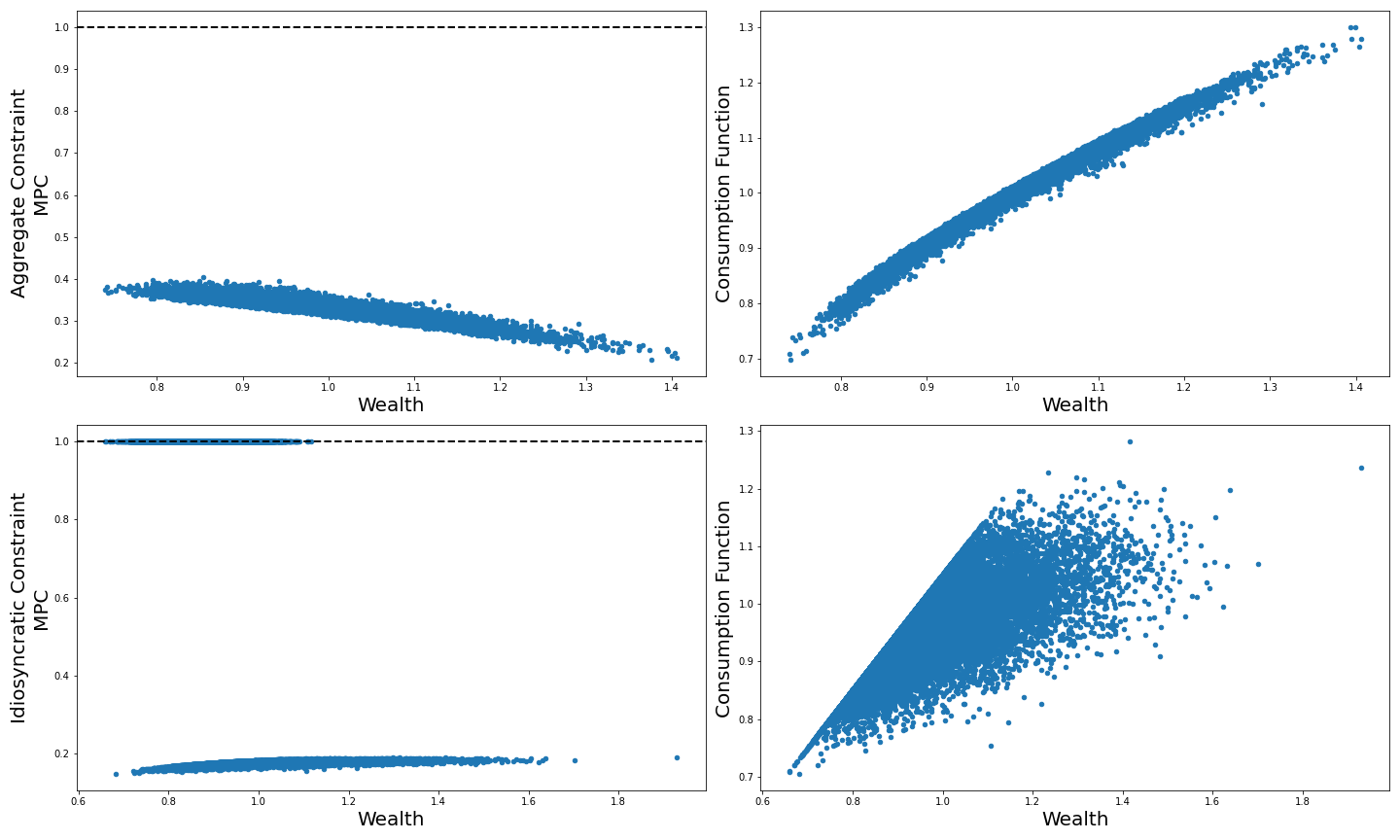}
	\caption{Immediate MPCs (left column) and consumption functions (right column) for the aggregate-constraint method (top row) and idiosyncratic-constraint method (bottom row), across all agents in a sample of states drawn from the ergodic distribution.}
	\label{agg_idio_mpc}
\end{figure}

Figure \ref{agg_idio_mpc} shows the consumption functions and MPCs generated by the intermediate constraint models. The since these fundamentally depict an idiosyncratic behaviour, it is not surprising that the aggregate-constraint version fails to capture important features such as the kinked consumption function and very high MPC for constrained agents, although we do see that MPC is decreasing in wealth, which is a feature that the soft-constraint model failed to capture. The takeaway here is that even though the aggregate-constraint version manages to perform relatively well quantitatively, it still fails to capture some of the key qualitative features of the model. The idiosyncratic-constraint (bottom row) does exhibit most of the features we expect to see, however, it is clear to see comparing to the hard-constraint version in Figure \ref{soft_vs_hard_mpc} that the scale is incorrect. The mean wealth is substantially lower than in the hard-constraint version, as a result of total output being below its equilibrium level. As a result too many agents hit their budget constraint, and therefore the consumption function is more aggressively kinked compared to the hard-constraint results.

\begin{figure}
    \centering
    \captionsetup{width=0.8\textwidth}
    \includegraphics[width=0.8\textwidth]{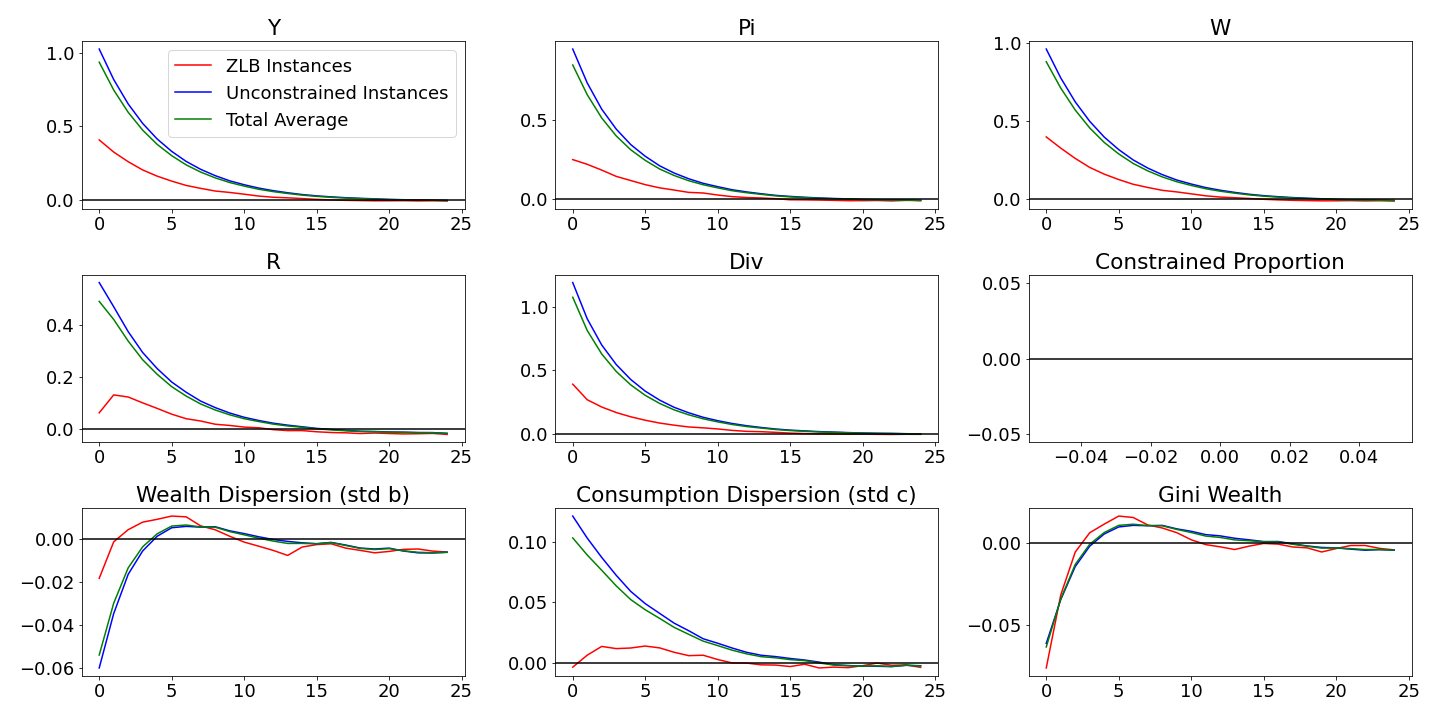}
    \includegraphics[width=0.8\linewidth]{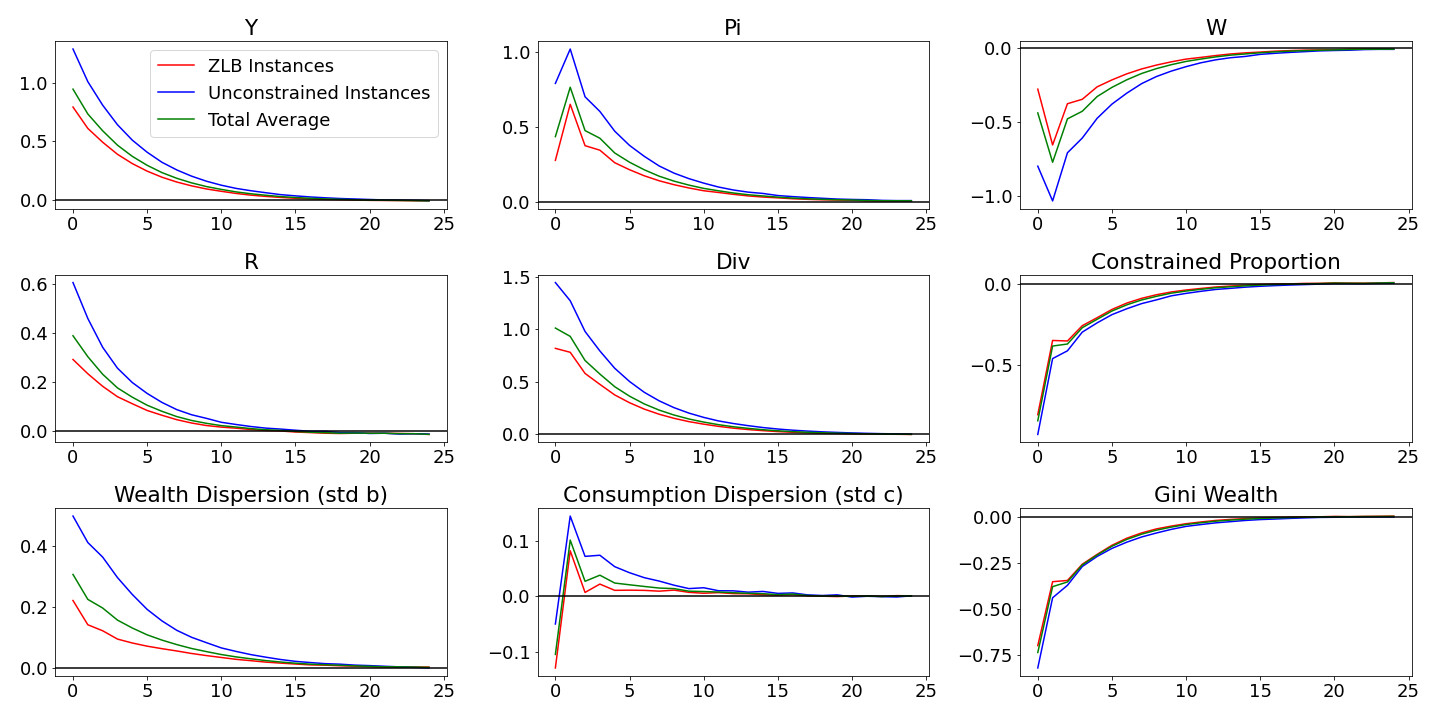}
	 \caption{Impulse Responses to a 2 standard deviation expansionary TFP shock ($\epsilon^A_t$) generated from the aggregate-constraint (top three rows) and idiosyncratic-constraint (bottom three rows) neural-networks.}
	\label{fig:agg_idio_irf}
\end{figure}

Figure \ref{fig:agg_idio_irf} shows the generalised IRFs for the aggregate and idiosyncratic constraint models. The idiosyncratic constraint model does not produce sensible impulse responses for most variables. On the other hand, the aggregate-constraint version is able to capture most of the movements in aggregate variables that the hard-constraint version does. However, consistent with previous findings, this version fails to capture the same dynamics related to cross-sectional distributions such as the Gini coefficient. In particular, as with the soft-constraint model since none of the agents are exactly at the budget constraint, we cannot construct a response for the constrained proportion. Neither of these intermediate models capture the dynamics of this model to the standard of the hard-constraint method. 

In practice, it is usually the case that while making one of the aggregate or idiosyncratic constraints of a heterogeneous agent model \textit{hard} is fairly trivial, applying both at the same time can be difficult. The results of this section seem to suggest that if all of the constraints cannot be made hard at the same time, then the largest benefit relative to the soft-constraint approach comes from satisfying the aggregate constraints by construction. However, these results also show that the hard-constraint approach is the only one that approximates the model to a satisfactory degree of precision and captures all of the important qualitative features of the economic model. Therefore, when solving DSGE models using neural-networks every effort should be made to make all of the constraints of the economic model hard, and this paper provides a methodology to achieve this for a large class of models.
 
\section{Conclusion} \label{sec:conclusion}

This paper has considered a detail of the application of neural-networks for solving DSGE models often not discussed in great detail, which is the implementation of economic constraints in the solution method. The results show that small differences in implementation, which may seem to be equivalent in theory may actually have very different implications in practice. This is primarily because every approximation technique has some degree of error, and when these errors relate to the constraints of a model they are more prone to propagation. Practitioners should be wary of how their treatment of constraints will impact the results obtained, and since there is still no disciplined way when working with neural-networks to determine how small the error generated by a solution has to be in order to be accepted, they should test their solutions carefully against expected quantitative and qualitative features of the model.

While no solution method is perfect, this paper has shown that the general approach that seems most appropriate is to give lexicographic priority to constraints --- that is to attempt to restrict the output space of the approximating function to the set of feasible outputs, and only then attempt to optimise relative to the first-order conditions. The approach commonly suggested by the existing literature, which is to use penalties to enforce constraints is theoretically sound and can work in some applications, but has some fundamental drawbacks that make it appealing to consider alternatives. This paper has provided a novel methodology that allows for all constraints to be satisfied by construction when using a neural-network as an approximating function in the important case of heterogeneous agents models with incomplete markets. The results demonstrate that this method is superior to alternative approaches which leave at least some of the constraints to be implemented via a penalty. Furthermore, the results consider intermediate cases in order to understand where this improvement comes from. These results seem to indicate that leaving any type of constraint to be soft, while simplifying the implementation, also decreases the quality of approximation substantially. Leaving aggregate constraints soft and idiosyncratic constraints hard makes it possible to capture the correct behaviour in a neighbourhood of the idiosyncratic constraint, but makes it difficult to correctly approximate aggregate variables and prices. The opposite was observed in the opposite case.

There are however some shortcomings of my suggested method that have already be identified. Therefore, there is a need for further research in order to improve these methods. In particular, there are corner cases where this method can fail. Hopefully further research into similar types of re-scaling will yield an equally effective method which is robust to these limitations. Furthermore, while I argue that the method I provide is applicable in a wide range of interesting applications, there is still a need to develop a more general solution method that can deal with general constraints.

\newpage

\printbibliography

\newpage

\appendix

\section{Neural-network Background} \label{neural_network_background}

A fully-connected, feed-forward neural-network is defined by \textit{layers} made by applying non-linear transformations known as \textit{activation functions} $\sigma_i$ to linear combinations of inputs \parencite{bengio2017deep}. The linear combinations are defined by multiplicative parameter matrices $W_i$ known as \textit{weights} and additive parameter matrices $b_i$ known as biases. In the case of the first layer the inputs are data, but otherwise the inputs are the outputs of the previous layer. Layers are then stacked a given number of times, until the output of the last layer is deemed to be the output of the neural-network. The dimensions of the output layer parameter matrix are chosen to match the desired output shape of the neural-network, and activation functions can in some cases be chosen in order to limit outputs to a certain range, depending on the application. For example, the activation function $\sigma(x) = exp(x)$ may be chosen if strictly positive outputs are desired.

A neural-network with $N$ layers consists of a set $\theta = \left\{ W_i, b_i \right\}_{i=1}^N$ of \textit{learnable parameters}. These are called \textit{learnable} to emphasise the difference with \textit{hyper-parameters}, such as the step size or learning rate. These hyper-parameters are usually chosen before the estimation procedure, and remain fixed throughout. The learnable parameters are usually initialised randomly, and then updated iteratively via SGD by a learning algorithm:

\begin{equation}
    \theta_{t+1} = \theta_t - \alpha L_\theta (X; \theta_t)
\end{equation}

where $\alpha$ is the \textit{learning rate}. In baseline SGD it is fixed, however, more advanced learning algorithms such as ADAM \parencite{kingma2014adam}, which is used in this paper, can be used to dynamically update the learning rate. Performing SGD requires calculating the gradient of the loss with respect to the learnable parameters. This can be done very efficiently with modern machine-learning software thanks to \textit{backpropogation}. Due to the nested nature of the neural-network, application of the chain rule implies continually recalculating the gradient of intermediate layers. By caching these intermediate gradients, this seemingly expensive computational operation can be performed rather quickly. The gradient can efficiently be calculated in this way without any user code by most modern machine-learning software, such as JAX, which was used in this project.

\section{Derivation of $\bar{z}$} \label{appendix:zbar}

We wish to ensure that $w^{i\star} < b^{i\star} \iff C\frac{z^{i\star}}{\sum_i z^i} < b^{i\star}$. Then for the max constrained element $i\star$ find $\bar{z}$ that is subtracted from $z^{i\star}$ to ensure this inequality holds. Set as an equality to find a lower bound for $\bar{z}$.

\begin{align}
    z^{i\star} - \bar{z} =& \frac{\sum_i\left(z^i - \bar{z}\right)b{i\star}}{C} \nonumber \\
    \implies \left(\frac{b^{i\star}}{C}L - 1\right)\bar{z} =& \frac{b^{i\star}\sum_iz^i}{C} - z^{i\star} \nonumber \\
    \implies \bar{z} =& \frac{\frac{b^{i\star}\sum_iz^i}{C} - z^{i\star}}{\frac{b^{i\star}}{C}L - 1}
\end{align}

\FloatBarrier
\section{Aggregate States and Prices Generated by Different Solution Methods}\label{appendix:agg_policies}

\begin{figure}[H]
    \centering
    \captionsetup{width=0.8\textwidth}
    \includegraphics[width=0.8\textwidth]{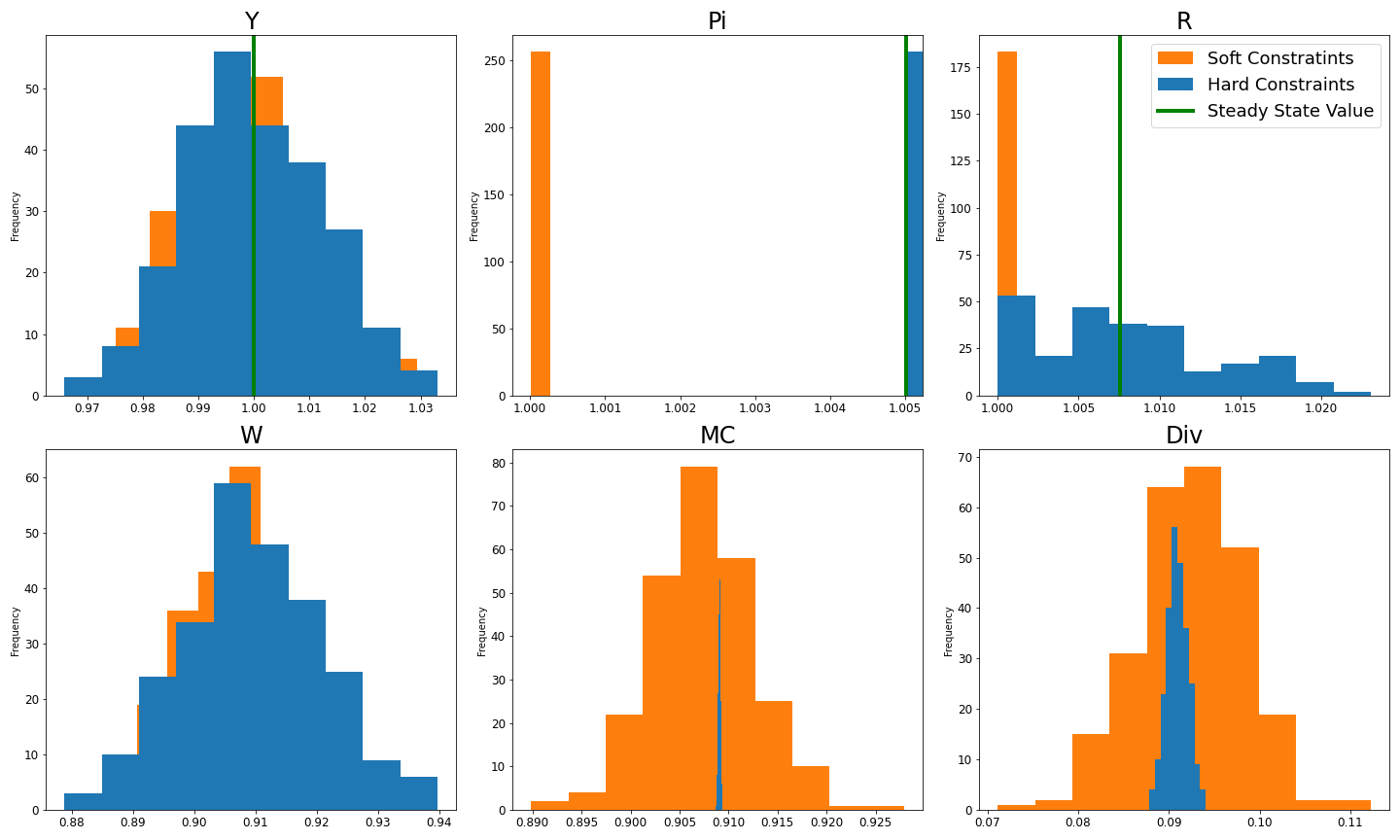}
	\caption{Distribution of aggregate variables and prices from a sample drawn from the ergodic distribution of sets for the hard-constraint method (Blue) and soft-constraint method (Orange). Steady-state values, where known, are marked as green vertical lines.}
	\label{fig:agg_dists_soft_hard}
\end{figure}

\begin{figure}[H]
    \centering
    \captionsetup{width=0.8\textwidth}
    \includegraphics[width=0.8\textwidth]{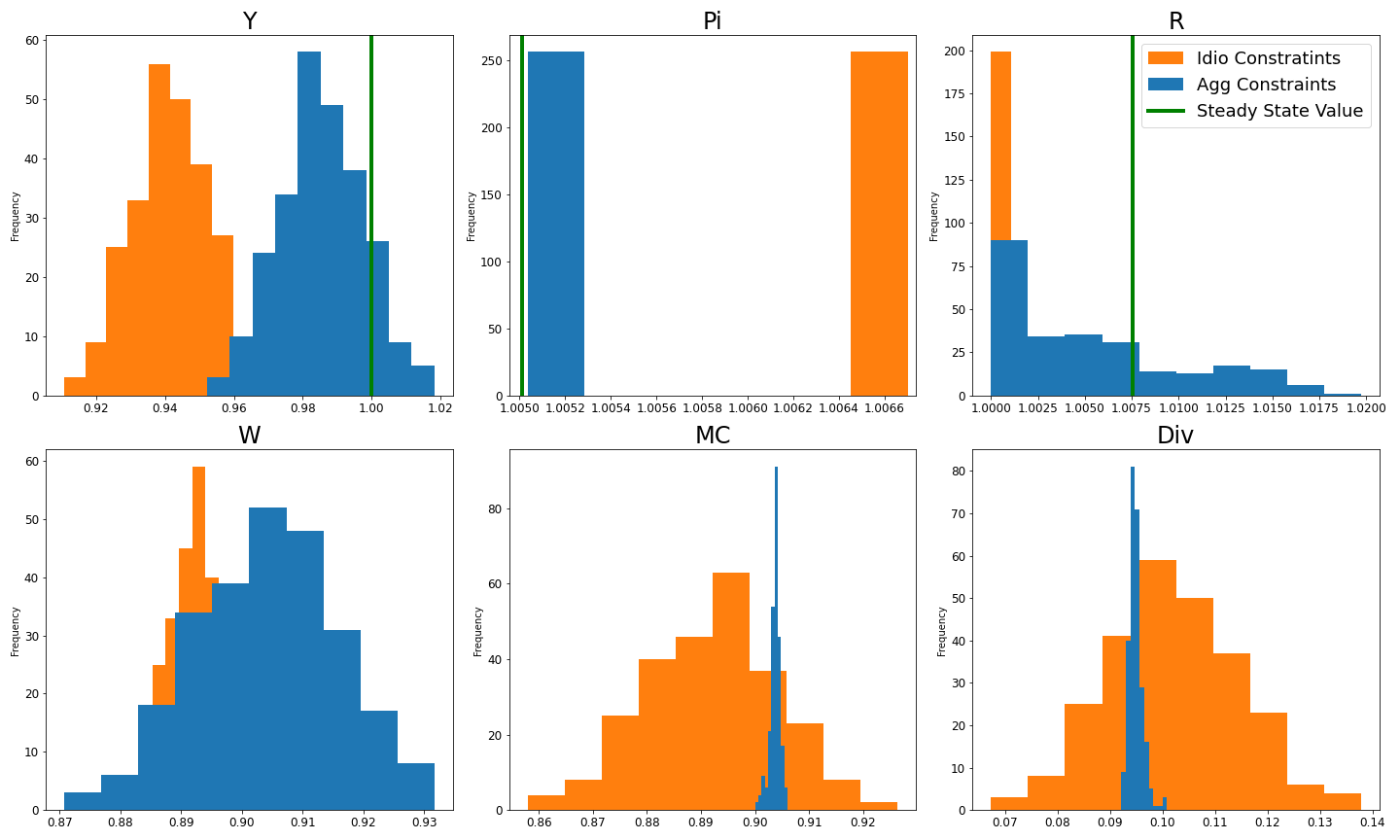}
	\caption{Distribution of aggregate variables and prices from a sample drawn from the ergodic distribution of sets for the aggregate constraint method (Blue) and idiosyncratic constraint method (Orange). Steady-state values, where known, are marked as green vertical lines.}
	\label{fig:agg_dists_agg_idio}
\end{figure}

\FloatBarrier
\newpage

\section{Neural-network Fitting Algorithm}

\begin{algorithm}
\caption{Neural-network Fitting Algorithm}
\label{alg:fitting}
\begin{algorithmic}
    \Require $\theta_0 \in \mathbb{R}^M, X_0 \in \mathbb{R}^{mb \times k}, I \in \mathbb{N}, N \in \mathbb{N}, tol_0 \in \mathbb{N}, tol_1 \in \mathbb{N}, tol_2 \in \mathbb{N}, \alpha \in \mathbb{R}_+$
    \Ensure $\theta_t \in \mathbb{R}^M, \theta_t \approx argmin_\theta \text{  } \mathbb{
    E}_{X,\Gamma} L(X, \theta; \Gamma)$
    \State $X \gets drawInitState()$
    \State $\Gamma \gets drawStructParams()$
    \State $\theta \gets \theta_0$
    \State $itsAtCurrentN \gets 0$
    \State $lastLoss \gets \infty$
    \State $n \gets 1$
    \State $i \gets 0$
    \While {$lastLoss > tol_0$ and $i < I$}
        \State $loss \gets \frac{1}{mb} \sum_{j=1}^{mb} L(X, \theta; \Gamma)$
        \State $\theta \gets \theta - \alpha \frac{1}{mb} \sum_{j=1}^{mb} L_\theta(X, \theta; \Gamma)$
        \If{$isnan(loss)$ or $loss(rc) > tol_1$}
            \If{$n > 1$}
                \State $n \gets n - 1$
            \EndIf
            \State $X \gets drawInitState()$
            \State $itsAtCurrentN \gets 0$
        \EndIf
        \If{$loss(rc) < tol_1$ and $\sim isnan(loss)$ and $itsAtCurrentN > tol_2$ and $n < N$}
            \State $n \gets n + 1$
            \State $itsAtCurrentN \gets 0$
        \EndIf
        \State $\Gamma \gets drawStructParams()$
        \For{$k \in \left\{ 1 ... n \right\}$}
            \State $X \gets S(X, \theta; \Gamma)$
        \EndFor
        \State $itsAtCurrentN \gets itsAtCurrentN + 1$
        \State $lastLoss \gets loss$
        \State $i \gets i + 1$
    \EndWhile
\end{algorithmic}
\end{algorithm}

\FloatBarrier
\newpage

\section{Soft-constraint Results for Different Penalty Weights}\label{appendix:soft_weights}

\begin{table}[H]
    \centering
    \begin{tabular}{|l|l|c|c|c|}
        \multicolumn{1}{c}{} & \multicolumn{1}{c}{} & \multicolumn{3}{c}{\small Penalty Weight} \\
        [-.3\normalbaselineskip] \multicolumn{1}{c}{} & \multicolumn{1}{c}{} & \multicolumn{3}{c}{\downbracefill} \\ \hline
        \bfseries Type & & \bfseries 1e1 & \bfseries 1e2 & \bfseries 1e4 \\
        & Total Loss & 2.46e-02 & 1.13e-02 & 3.39e-02 \\ \hline
        & Euler Equation Loss & 2.39e-03 & 5.64e-03 & 3.21e-03  \\ 
        Optimality & Phillips Curve Loss & 2.09e-02 & 5.14e-03 & 3.06e-02  \\ 
        & Labour Supply Loss & 1.09e-32 & 1.10e-32 & 1.09e-32 \\ \hline
        & KKT Loss & 1.93e-04 & 2.74e-04 & 4.56e-06 \\ 
        Constraints & Output Constraint Loss  & 1.26e-04 & 1.03e-04 & 5.06e-05 \\ 
        & Net Supply of Bonds & 9.24e-04 & 1.56e-04 & 6.05e-05 \\ \hline
    \end{tabular} 
    \caption{Average total loss and its constituent components evaluated over the last 50 iterations of the fitting procedure for the models fit using the soft-constraint version with various weights applied to the constraint related penalties. All runs were warm-started with parameters from the baseline configuration presented for the soft-constraint methodology Section \ref{results}.}
    \label{tab:soft_weights_losses}  
\end{table}    

\FloatBarrier
\newpage

\section{Settings Used for Generating Results}

\begin{table}[H]
    \centering
    \begin{tabular}{|l|c|c|}
        \hline
        \bfseries & \bfseries Value (Soft-Constraint & \bfseries Value (Hard-Constraint \\
        \bfseries Parameter & \bfseries and Idio-Constraint) & \bfseries and Agg-Constraint) \\ \hline
        Training Iterations & 200000 & 100000 \\ \hline
        Batch Size & 256 & 256 \\ \hline
        Max Forward Sims per Update & 20 & 20 \\ \hline
        Learning Rate & 1e-6 & 1e-4 \\ \hline
        Precision & x64 & x64 \\ \hline
        Scale of Initial Parameters & 1e-2 & 1e-2 \\ \hline
        eps (parameter for ADAM optimiser) & 1e-12 & 1e-12 \\ \hline
    \end{tabular} 
    \caption{Hyper-parameters used while training models to generate results presented. The soft-constraint version was allowed to train longer at a lower learning rate compared to the hard-constraint version because it requires significantly more precision in parameter updates to converge around the constraints. The hard-constraint version on the other hand can be used with a larger learning rate (and thus, less precise parameter updates), which allows it to converge more quickly while maintaining its inherently high level of precision.}
    \label{tab:settings}  
\end{table}    

\end{document}